\documentclass[pre,twocolumn,preprintnumbers,amsmath,amssymb,floatfix]{revtex4}

\usepackage{graphicx}
\usepackage{dcolumn}
\usepackage{bm}

\usepackage{hyperref}

\usepackage{enumerate}

\usepackage{xcolor}

\newcommand{\nn}{\nonumber }

\newcommand{\rr}{{\mathbf r}}
\newcommand{\kk}{{\mathbf k}}
\newcommand{\f}{{\mathbf f}}

\newcommand{\p}{\partial}

\newcommand{\BEQ}{\begin{equation}}
\newcommand{\EEQ}{\end{equation}}
\newcommand{\BEA}{\begin{eqnarray}}
\newcommand{\EEA}{\end{eqnarray}}

\newcommand{\pv}{\mathbf{p}}

\newcommand{\ii}{\mathbf{i}}

\newcommand{\xx}{\mathbf{x}}
\newcommand{\yy}{\mathbf{y}}
\newcommand{\zz}{\mathbf{z}}

\newcommand{\eeta}{\boldsymbol{\eta}}

\begin{document}



\title{Statistical Field Theory and Effective Action Method for scalar Active Matter}

\author{M. Paoluzzi$^{1,2}$}
 \email{Matteo.Paoluzzi@roma1.infn.it}

%
\author{C. Maggi$^{3,2}$}
%

\author{A. Crisanti$^{2,1}$}

\affiliation{
$^1$ ISC-CNR,  Institute  for  Complex  Systems,  Piazzale  A.  Moro  2,  I-00185  Rome,  Italy \\
$^2$ Dipartimento di Fisica, Sapienza University of Rome, Piazzale  A.  Moro  2, I-00185, Rome, Italy \\
$^3$ Institute of Nanotechnology (CNR-NANOTEC), Soft and Living Matter Laboratory, Rome, Italy
}



\date{\today}

%

\begin{abstract}
We employ Statistical Field Theory techniques for coarse-graining the steady-state properties of Active Ornstein-Uhlenbeck particles.
The computation is carried on in the 
framework of the Unified Colored Noise approximation that allows an effective equilibrium picture.
We thus develop a mean-field theory that allows to describe in a unified framework the phenomenology of scalar Active Matter. 
 In particular, we are able to describe through spontaneous symmetry breaking mechanism two peculiar features of
 Active Systems that are (i) The accumulation of active particles at the boundaries of a confining container, and
(ii) Motility-Induced Phase Separation (MIPS). 
\textcolor{black}{We develop a mean-field theory for steric interacting active particles undergoing to MIPS and for Active Lennard-Jones (ALJ) fluids.}
\textcolor{black}{Within this framework}, we 
discuss the universality class of MIPS and ALJ \textcolor{black}{showing that it falls into Ising universality class.}
We \textcolor{black}{thus} compute analytically the critical line $T_c(\tau)$ for both models. In the case of MIPS, $T_c(\tau)$ gives rise to a
reentrant phase diagram compatible with an inverse transition from liquid to gas as the strength of the noise decreases.
\textcolor{black}{However, in the case of particles interacting through anisotropic potentials, }
the field theory acquires a $\varphi^3$ term that, \textcolor{black}{in general, cannot be canceled performing the expansion around the critical point.}
In this case, the \textcolor{black}{Ising} critical point might \textcolor{black}{be replaced}
by a first-order phase transition \textcolor{black}{region}. 
\end{abstract}
\maketitle

\section{Introduction}

In nature, there are many and diverse examples of Living Materials \cite{klopper2018physics} 
ranging from epithelial monolayers \cite{trepat2018mesoscale}, bacterial colonies \cite{zhang2010collective}, or dense drops of ants \cite{feinerman2018physics}. 
Even though the elementary units composing such materials are complex biological objects, many of the emerging 
collective behaviors can be described using concepts of Condensed Matter and, in particular, 
through the analytical and numerical tools developed during the last decades in Active Matter \cite{Bechinger17,Marchetti13,ramaswamy2010mechanics,cates2012diffusive,Cavagna14}. 
Active Systems  are defined as a class of nonequilibrium systems consisting of interacting 
entities that individually dissipate energy to generate forces and motion and exhibit self-organized behavior at large scales.

Active systems can develop complex patterns that change dynamically, as in the case of flocking \cite{Cavagna14}. 
Flocking has been observed at very different scales ranging from birds \cite{ballerini2008interaction} to epithelial monolayers \cite{malinverno2017endocytic}. However, Active
Systems can also give rise to pattern formation and condensed phases with structural properties remarkably similar
to those of ordinary materials, e. g., gas-liquid phase transition \cite{bialke2015active,bialke2013microscopic,bialke2012crystallization,stenhammar2014phase}, glassy or jamming states \cite{Szamel15,Fily12,Berthier13}, polar order \cite{liverpool2003instabilities}, and nematic order \cite{keber2014topology}.
Pattern formation in Active Matter is driven by  out-of-equilibrium dynamics and thus 
these condensed phases are emergent properties of steady-state configurations
that, in general, cannot be described by a Boltzmann distribution.
In particular, the dependency on the microscopic dynamics makes 
hard to establish possible universality classes. 
Understanding how condensed phases in Active Systems are related to those in equilibrium plays an important role in both,
basic and applied science. In basic science, it would allow to gain insight into 
the concept of universality in non-equilibrium systems.
In applied science, for instance, it would allow to design living and synthetic materials with desired structural properties. 

In this paper, we present a study on universal properties of a specific class of Active Matter system that
is described on large-scale by a scalar field theory. 
We start from a microscopic model where particles are self-propelled  
through a persistent noise \cite{Maggi,Szamel14,Koumakis14} and we perform a coarse-graining 
using the machinery of Statistical Field Theory. We show that,
within the framework of Unified Colored Noise (UCN) approximation \cite{Jung87,Hanggi95},
the structural properties of the system can be described through an opportune Effective Action \cite{qft-ne}. 
Active field theories based on the dynamical evolution of opportune set of order parameters 
have been largely employed for capturing the large scale behavior of active systems \cite{wittkowski2014scalar,nardini2017entropy,caballero2018bulk,tiribocchi2015active,stenhammar2013continuum,speck2014effective,menzel2016dynamical}. 
In our work, we show that some 
peculiar behaviors of Active Systems can
be captured through an Equilibrium Statistical Field Theory approach.

\subsection{Summary of Results}
We aim to develop a field theoretical description of Active Systems using the machinery of equilibrium Statistical Physics.
As a main result, we obtain that a scalar Active System can be described by an Effective Action that counts of non-local terms.
The Effective Action is a functional of the scalar order parameter $\varphi$.
These non-local terms can be systematically studied at mean-field level.

Focusing our attention on mean-field computations, we show that: 
\begin{enumerate}[(i)]
\item The Effective Action Method \textcolor{black}{reproduces some basic}
properties of Scalar Active Matter.
\item We can put into a unified theoretical framework two peculiar phenomena of Active Matter: the accumulation of active particles at the boundaries of a container, and Motility-Induced
Phase Separation (MIPS) \cite{Tailleur08}.  
In particular, both phenomena can be interpreted in term of a spontaneous symmetry breaking \textcolor{black}{of} $\varphi \to - \varphi$ \textcolor{black}{symmetry}.  
\item We can discuss in a simple picture the universality class of MIPS and Active Lennard-Jones Fluids (ALJ).
In particular, in both cases, we can compute analytically the critical line $T_c(\tau)$. Moreover, the mean-field theory suggests that both MIPS, and ALJ fall into Ising universality class. 
\item In the case of MIPS, the curve $T_c(\tau)$ develops a reentrance in the phase diagram indicating
that the system undergoes an {\it inverse transition} from liquid to gas as the effective temperature $T$ is decreased above a threshold value $\tau_{th}$. This means that the condensed MIPS phase evaporates as the effective temperature decreases.
\item We show that\textcolor{black}{, close to the critical point,} the Effective Action explicitly breaks
the symmetry $\varphi \to - \varphi$ through a $\varphi^3$ terms that vanish for spherical active particles and it is non zero for 
\textcolor{black}{anisotropic pair potentials, i. e.,}
rod-shaped swimmers. Moreover, \textcolor{black}{although this term does not touch the Ising universality class of the critical point,} 
the presence of this term suggests that the MIPS critical point can be hidden by a first-order phase transition.
\end{enumerate}

The paper is organized as follows. In Sec (\ref{setup}) we introduce the theoretical framework. 
In Sec. (\ref{micromodel}) we \textcolor{black}{define} the microscopical model.
In Sec. (\ref{one-body}) we employ the theoretical set-up for studying one-body \textcolor{black}{interactions} in Active Matter. In particular, we show how 
the accumulation of active particles at the boundaries of a container can be interpreted as a spontaneous symmetry breaking in the
Effective Action. In Sec. (\ref{many-body-theo}) we address many-body \textcolor{black}{interactions} and discuss the general features of the theory. 
In Sec. (\ref{mean-field-theo}) we discuss the mean-field \textcolor{black}{approximation}. In particular, we study the mean-field phase diagram 
of purely repulsive potentials and Lennard-Jones potentials. In sec. (\ref{mips-phi3}) we discuss the effect of anisotropic interaction on MIPS.
Finally, in Sec. (\ref{sec-disc}) we present our conclusions.

\section{Theoretical set-up} \label{setup}
We start our discussion with introducing a formalism in equilibrium Statistical Mechanics that allows to perform the
coarse-graining of generic $n-$body interactions.  
We consider a system composed by $N$ particles in $d$ spatial dimensions confined in a box of side $L$
and volume $V=L^d$. To keep the presentation simple, 
we indicate with $[r_i]$ or $(r_i)$ a generic particle configuration $(r_1,...,r_N)$ where $r_i$ is a
$d-$dimensional vector representing the position of the particle $i$. 
For the sake of completeness, we consider a hamiltonian system composed of classical particles whose degrees of freedom 
are canonical coordinates and conjugated momenta. In the next section,  we will apply the formalism for computing 
configurational integrals in the case  Active Systems, and thus we will \textcolor{black}{neglect} generalized momenta.
 
Denoting $p_i$ the momentum
of the particle $i$, we assume that the mechanical properties of the system are fully specified through the 
hamiltonian function $H[p,r]$ that is
\BEQ 
H[p_i,r_i] = \sum_{i=1}^N \frac{p_i^2}{2 m} + \mathcal{H}[r]
\EEQ 
where the configurational part $\mathcal{H}[r]$ takes into account $1-$body, $2-$body and $k-$body
interactions, with $k\geq 3$. \textcolor{black}{The configurational energy is}
\BEA \label{nbody}
\mathcal{H}[r_i] &=& \sum_{i=1}^N \phi_1(r_i) + \frac{1}{2} \sum_{i,j}^{1,N} \phi_2(r_i,r_j) + \\ \nn
&+& \sum_{k \geq 3}^N \phi_3(r_i,...,r_k) \; .
\EEA 
Tthermodynamics is obtained through the computation of the partition function $Z$ \cite{Plischke} that is
\BEA \label{thermo}
Z &=& \int \prod_i \frac{dr_i}{\lambda^N} \, e^{-\beta H[r_i] + \beta \mu N}  \\ \nn
f(\beta)& =& -\lim_{N,V \to \infty } \frac{1}{\beta V} \ln Z
\EEA 
with $f(\beta)$ the density of free energy. In Eq. (\ref{thermo}), the thermodynamic limit $N,V\to \infty$
is performed maintaining fixed the mean density $\rho = N/V$.
We have introduced the thermal wavelength $\lambda = \sqrt{\frac{h^2}{2 \pi m k_B T}} $. $\beta$ the
inverse temperature, i.e, $\beta=1/k_B T$, $h$ is the Planck constant, and $\mu$ is
the chemical potential. Working in 
natural unit, one has $k_B=1=h$ and thus $\beta = T^{-1}$.

For studying the behavior of the system on large scales, we perform a coarse-graining based on the local density field
$\psi(r)$ that is
\BEQ
\psi(r) = \sum_i \delta (r - r_i) \, .
\EEQ
Using standard manipulations \cite{qft-ne,ZinnJustin}, one can enforce the field $\psi(r)$ into Eqs. (\ref{thermo}) through a delta functional 
\BEQ \label{delta_func}
\int \mathcal{D} \psi(r) \, \delta \left[ \psi(r) - \sum_i \delta(r - r_i) \right] = 1
\EEQ
and thus we can write
\BEQ
Z=\int \mathcal{D} \psi(r) \mathcal{D} \hat{\psi}(r) \, e^{-G[\hat{\psi} , \psi]}
\EEQ
where the auxiliary field $\hat{\psi}$ has been introduced for representing the delta functional introduced in Eq. (\ref{delta_func}).
The details of the computation are provided in the appendix (\ref{theory}).
The functional $G$ takes the form
\BEA
-G &\equiv& -S[\psi] + \int dr \, \left[ \hat{\psi}(r) - b(r) \right] \psi(r)  + N \ln z  \\ \nn
z &\equiv& \int \frac{dr}{\lambda} e^{-\hat{\psi} (r)} \, ,
\EEA
with
\BEA
b(r) &\equiv& \beta \phi_1 (r) - \beta \mu \\ \nn 
S[\psi] &\equiv& \frac{1}{2}\int dr dr^\prime \, \psi(r) \Delta^{-1} (r,r^\prime) \psi(r^\prime) + V[\psi] \\ \nn 
\Delta^{-1}  &\equiv& \beta \phi_2 (r,r^\prime ) \\ \nn
V &\equiv& \sum_{k \geq 3 } \frac{1}{k !} \int dr_1 ... dr_k \beta \phi_k (r_1,...,r_k) \psi(r_1) ... \psi(r_k) \; .
\EEA 
Performing a shift to the field $\hat{\psi} - b \to \hat{\psi}$, the thermodynamics can be recasted in the following
form
\BEA \label{new_gen}
e^{W[\hat{\psi}]} &\equiv& \mathcal{N}^{-1} \int \mathcal{D} \psi e^{-S[\psi] + \int dr \, \hat{\psi}(r) \psi(r)} \\ \nn
Z &=& \mathcal{N} \int \mathcal{D} \hat{\psi} \mathcal{D} \psi \, e^{W[\hat{\psi}] + N \ln \int \frac{dr}{\lambda} e^{\hat{\psi}(r) - b(r)}} \\ \nn
\mathcal{N} &\equiv& \int \mathcal{D} \psi\, e^{-S[\psi]} \; .
\EEA
As one can appreciate, the auxiliary field $\hat{\psi}$ in Eq. (\ref{new_gen}) plays the role of external source in a quantum field theory \cite{ZinnJustin}. 
It is worth noting that $W[\hat{\psi}] \textcolor{black}{\sim} O(N)$. 
According to the definition of $W[\hat{\psi}]$, the $n-$point correlation function of the theory can
be generated through functional differentiation. For instance, we have
\BEA \label{ave_field} 
\frac{\delta W}{\delta \hat{\psi}(r)} &=& \langle \psi(r) \rangle \\ \nn
\frac{\delta^2 W}{\delta \hat{\psi}(r) \delta \hat{\psi}(s)} &=& \langle \psi(r) \psi(s) \rangle_C 
\EEA 
where the average is defined as follows 
\BEQ 
\langle \mathcal{O} \rangle \equiv \frac{\int \mathcal{D} \psi \, e^{-S[\psi] + \int dr\, \psi(r) \hat{\psi} (r) } \mathcal{O}}{\int \mathcal{D} \psi \, e^{-S[\psi] + \int dr\, \psi(r) \hat{\psi} (r) }}
\EEQ 
and $\langle \dots \rangle_C$ indicates a connected correlation function.

\subsection{Stationary points, mean-field approximation and fluctuations around the mean-field solution}
It is well known that mean-field theories neglect fluctuations \cite{Parisi_stat}. In particular, in a mean-field approximation,
one usually replaces the value of the order parameter in a given point of the space with its mean-value in the same point, i.e., 
$\psi(r) \to \langle \psi(r) \rangle$ and $\langle \psi(r) \psi(r^\prime) \rangle \to \langle \psi(r) \rangle \langle \psi(r^\prime) \rangle$.
The latter replacement holds whenever $\langle \left[ \psi(r) -   \langle \psi(r) \right]^2 \rangle  \sim V^{-1/2}  \sim N^{-1/2}$ that
vanishes in the thermodynamic limit.
Through Eqs. (\ref{new_gen}) we define the functional $F[\hat{\psi}]$ that is
\BEQ 
F[\hat{\psi} ] = W[\hat{\psi}] + N \ln \int \frac{dr}{\lambda} e^{-\hat{\psi}(r) + b(r)} \, .
\EEQ 
Since the exponent in the functional integral 
is of order $N$, it makes
sense to perform a saddle-point approximation for evaluating \textcolor{black}{Z} and then 
compute systematically the stability of the stationary point $\hat{\psi}_{SP}$ against fluctuations. 
Considering field configurations $\hat{\psi} = \hat{\psi}_{SP} + \Delta \hat{\psi}$, we can thus
write
\BEA \label{SP_FLUCT}
F[\hat{\psi} ] &=& F[ \hat{\psi}_{SP} ]  + F[ \Delta \hat{\psi} ] \\ \nn
F[ \Delta \hat{\psi}]  &\equiv& \frac{1}{2} \int dr ds\, \Delta \hat{\psi}(r) \mathcal{G}(r,s) \Delta \hat{\psi}(s)
\EEA
where the kernel in Eqs. (\ref{SP_FLUCT}) is
\BEQ
\mathcal{G}(r,s) = \left. \frac{\delta^2 F}{\delta \hat{\psi}(r) \delta \hat{\psi}(s) } \right|_{SP}
\EEQ
and we have used the fact that 
\BEQ \label{gen_th_sp}
\left. \frac{\delta F}{\delta \hat{\psi} (r)} \right|_{SP} = 0
\EEQ
In term of the generating functional $W$, we can define
\BEA \label{saddle_gen_theory}
\rho(r) &=&  \left. \frac{\delta    W }{\delta \hat{\psi} (r)} \right|_{SP} \,  \\ \nn 
G (r,s) &=&  \left. \frac{\delta^2 W}{\delta \hat{\psi} (r)  \delta \hat{\psi} (s) } \right|_{SP}  \,.
\EEA
Neglecting terms that scales with $1/N$, we can rewrite the partition function that turns to be factorized as follows
\BEA \label{fluct_sp}
Z &=& Z_{SP} \, Z_{Fluct} \\ \nn
\ln Z_{SP} &\equiv&  -F[ \hat{\psi}_{SP}]  \\ \nn
Z_{Fluct} &\equiv&  \int \mathcal{D} \varphi \, e^{- \int dr ds \,  \varphi(r) \mathcal{G}(r,s)  \varphi(s) } \\ \nn
\mathcal{G}(r,s) &=& G^{-1}(r,s) - \int dz \, G^{-1} (z,r) \rho(z) G^{-1} (z,s)
\EEA
where we have performed the change of variable $\Delta \hat{\psi} (r) = \ii \int ds \, G^{-1}(r,s) \varphi(s)$ (computation 
details are provided in appendix (\ref{fluctua})). In Eqs. (\ref{fluct_sp}), $\mathcal{G}$ is the full propagator of the theory that takes contribution from
both, the two-body potential $\phi_2$ and the $k-$body potentials $\phi_k$.

Using the saddle-point relation Eqs. (\ref{gen_th_sp}), we can write 
\BEQ
\rho(r)=\frac{N \, e^{- \hat{\psi}(r) + b(r)}}{z}
\EEQ
obtaining in this
way the following expression for the thermodynamics in the mean-field approximation
\BEA
Z_{SP} &=& e^{-F[\rho(r)]} \\ \nn
f_{SP}(\beta) & =& \lim_{V,N \to \infty } \frac{ F[\rho(r)] }{\beta V} \; .
\EEA


\section{Microscopic model of Active Particles} \label{micromodel}
As a microscopic model, we consider active particles that are self-propelled through
Ornstein-Uhlenbeck processes, i. e., 
Active Ornstein-Uhlenbeck particles (AOUPs) \cite{Maggi,Koumakis14,Farange15,Marconi1,Marconi2,Marconi3}.

The equations of motion for AOUPs are \cite{Szamel14,Maggi,Farange15,Szamel15,Szamel16}
\BEA \label{micro_mo}
\dot{\xx}_i          &=& \mu (\f_i^a + \f_i^{m} + \f_i^{ext}) \\ \nn
\tau \,\dot{\f}_i^a &=& -\f_i^a + \eeta_i \; .
\EEA
Here we indicate with $\xx_i$ a vector in $d$ spatial dimensions that determines the position of the
active particle $i$. The self-propulsive force is $\f_i^a$ \textcolor{black}{characterized by a persistence time $\tau$.} 
The total mechanical force acting on the particle $i$ is $\f_i^{m}$. The term $\f_i^{ext}$ represents forces due to
external conservative fields. Finally, $\mu$ is the mobility.
The noise $\eeta_i $ satisfies $\langle \eta_i^\alpha \rangle = 0$ and $\langle \eta_i^\alpha (t) \eta_j^{\beta} (s) = 2 T \mu \delta_{ij} \delta^{\alpha \beta} \delta(t-s) $,
where greek symbols indicate cartesian components. According to Eqs. (\ref{micro_mo}), the self-propulsion is exponentially correlated in time,
in particular one has $\langle f_i^{a,\alpha}(t) f_j^{a,\beta}(s) \rangle = \frac{2 T \mu }{\tau}e^{- | t - s | / \tau }$ \cite{Fily12}.

Let us introduce the steady-state distribution $P_{ss}(\xx_i) \equiv P_{ss}(\xx_1,...,\xx_N) = \lim_{t \to \infty} P_{ss}(\xx_1(t),...,\xx_N(t))$.
 Adopting the Unified Colored Noise approximation \cite{Jung87,Hanggi95},
it has been shown \cite{Maggi,Marconi1,Marconi2,Marconi3,Marconi17} that the approximate solution for $P_{ss}$ takes the form
\BEA \label{ucn_many}
P_{ss}(\xx_i) &=& Z_\beta^{-1} \exp{ \left( -\beta H_{UCN} [\xx_i] \right) } \\ \nn
H_{UCN}[\xx_i] &=& H_0[\xx_i] + H_1[\xx_i] + H_2[\xx_i] \\ \nn
H_0[\xx_i] &\equiv& \frac{1}{2} \sum_{i,j} \phi_2(\xx_i, \xx_j) + \sum_{i} \phi_1(\xx_i)\\ \nn
H_1[\xx_i] &\equiv& \frac{\tau}{2} \sum_i \left( \nabla_{\xx_i} H_0 \right)^2 \\ \nn
H_2[\xx_i] &\equiv& - \beta^{-1} \ln \det \mathbf{M} \\ \nn
\mathbf{M} &\equiv& M_{ij}^{\alpha \gamma}= \delta_{ij}^{\alpha \gamma} + \tau \frac{\p^2 H_0}{\p x_i^{\alpha} \p x_j^{\gamma}} \; ,
\EEA
with $Z_\beta$ fixed by the normalization condition $\int \prod_i d\xx_i P_{ss}(\xx_i ) = 1$.
The hamiltonian $H_0$ is responsible for both, the mechanical interactions and the interactions with
external fields, i. e., $\f_i^{m} + \f_i^{ext} = -\nabla_{\xx_i} H_0[\xx_i]$.
As one can see, $P_{ss}(\xx_i)$ takes the form of an equilibrium distribution where the hamiltonian $H_0$
is replaced by an effective one that we named $H_{UCN}[\xx_i]$. In this way, the structural properties of the system on
large-scale can be computed  
through an equilibrium Statistical Mechanics theory based on $H_{UCN}[\xx_i]$.
In the present paper, we will apply the theoretical machinery introduced 
in the previous section for coarse-graining the equilibrium-like model \cite{PaoluzziXY}.   

The effective hamiltonian  $H_{UCN}[\xx_i]$ is composed of three contributions. The term $H_0$
is the mechanical energy of the system in equilibrium. The terms $H_1$ and $H_2$ introduce non-local
many-body interactions among particles that disappear in the limit $\tau \to 0$. The computation of $\det M$
in $d$ spatial dimensions requires the diagonalization of a $dN \times dN$ matrix. Moreover, Eq. (\ref{ucn_many})
requires that $\mathbf{M}$ must be a positive definite matrix. Here we will consider a small $\tau$ expansion 
based on the approximation $\det( \delta_{ij}^{\alpha\gamma} + \tau D_{ij}^{\alpha\gamma} ) = 1 + \tau Tr D_{ij}^{\alpha\gamma} + o(\tau^2)$ (see appendix \ref{ucn_coarse} ),
with $D_{ij}^{\alpha\gamma}$ the Hessian matrix. It is worth noting that the small $\tau$ breaks when the hessian develops 
negative eigenvalues of order $\tau^{-1}$. In this work, we will restrict our computation in cases where $D_{ij}^{\alpha\gamma}$ is  positive definite.
 %

\section{One-body interactions} \label{one-body}
Since active particles break the fluctuation-dissipation theorem at single 
particle level \cite{maggi2017memory,chen2007fluctuations,Maggi14}, they show 
intriguing non-equilibrium phenomena even at the level of gas of non-interacting particles.
In particular, when active particles are confined by a container or immersed into a confining potential, i. e.,
a central field that tends to confine particles into a region of space,
the steady-state distribution strongly deviates from the Boltzmann distribution showing a double-peaked structure
at high persistence time \cite{das2018confined,caprini2019activity,Bechinger17,sevilla2019stationary}. The double-peaks signal
the accumulation of active particles at the boundary of the confining potential instead in the
center, i. e., where the potential is zero. This effect is due to the fact that active particles remain trapped in regions of space where the external field exerts 
forces that balance the self-propulsion force rather than in a region where the potential is zero.
Accumulation at the boundaries has been also observed in experiments \cite{VladescuPRL2014}. 

 In this section, we will see that this non-equilibrium \textcolor{black}{condensation} phenomenon can be interpreted as a spontaneous
symmetry breaking \textcolor{black}{in the effective hamiltonian $H_{eff}(x)$. For $\tau=0$, one recovers the usual equilibrium phenomenology.}
In particular, while a system in equilibrium develops condensation at the bottom of the external
potential, \textcolor{black}{i. e., where $H^\prime(x_0=0)=0$, }
active particles condensate away from the bottom \textcolor{black}{in regions where $H_{eff}^\prime(x_0)=0$.} 
This can be rationalized through our formalism in term of an effective 
potential that develops a double-well structure as soon as $\tau > 0$. The condensation away from the center of the trapping potential
can be thus interpreted as a spontaneous symmetry breaking in an effective equilibrium picture.
As a benchmark for the formalism, in appendix 
Eq. \ref{1dgas} we report the case of gas in equilibrium in one spatial dimension embedded into an external potential
where no accumulation  at the boundaries take place and localization appears at the bottom of the potential.

%
%

\subsection{Gas of Active Particles in external potentials}
We consider a gas of active particles embedded into an external potential $A(x)$ in
one spatial dimension.
%
As we showed in the previous section, the steady-state properties of the system can be
obtained through the computation of the following partition function
\BEQ
Z_{\beta} = \int \prod_i dx_i \, e^{-\beta H_{eff}[x_i]}
\EEQ
where the effective hamiltonian $H_{eff}[x_i]$ is
\BEA
H_{eff}[x_i] &=& \sum_{i=1}^N A(x_i) + \frac{\tau}{2} \sum_{i=1}^N (A^\prime (x_i))^2  + \\ \nn
&-&  \beta^{-1} \sum_{i=1}^N \ln \left[ 1 + \tau A^{\prime\prime} (x_i)\right]  
\EEA
we can thus write
\BEA
H_{eff}[x_i] &=& \sum_{i=1}^N B(x_i) \\ \nn
B(x_i) &\equiv& A(x_i) + \frac{\tau}{2} A^\prime(x_i)^2 - \beta^{-1} \ln \left( 1 + A^{\prime\prime} (x_i) \right) 
\EEA
In terms of the density field $\psi(x)=\sum_i \delta(x - x_i) $, the partition function is
\BEA
Z_\beta &=& \int \mathcal{D} \psi(x) \, e^{-\beta G[\psi]} \\ \nn
G[\psi] &\equiv& \int dx \, \psi(x) \left[ \beta B(x) + \ln \psi(x) \right] \; .
\EEA
We can approximate the integral using the saddle-point approximation
\BEQ 
Z_\beta \sim e^{-\beta G[n]}
\EEQ 
where we have indicated with $n(x)$ the field configuration that solves
\BEQ 
\left. \frac{\delta G}{\delta \psi(x)} \right|_{\psi = n} = 0 \; .
\EEQ 
The steady-state density profile reads
\BEQ
n(x) = c \, e^{-\beta B(x)} \; .
\EEQ
At small temperatures the density profile is dominated by the minima $x_0$ of $B(x)$ and thus $n(x)$
is concentrated around \textcolor{black}{those} minima, i. e., $n(x)\sim \sum_{\alpha} \delta(x - x_0^{\alpha})$, where
$\alpha$ labelled \textcolor{black}{the solutions of}
\BEQ
\left.\frac{dB}{dx} \right|_{x=x_0^\alpha} = 0 \; , \left.\frac{d^2 B}{dx^2} \right|_{x=x_0^\alpha} \geq 0 \; .
\EEQ
We specialize our computation in the case of harmonic and anharmonic trapping, the latter due to a quartic confining potential, i. e., $\alpha=2$. 
\subsection{Soft confining potentials in Active Matter}
Self-propulsion naturally introduces a typical scale for the forces, i. e., the self-propulsion force $f_s$, 
that competes with the other force fields interacting with the active particles. Due to this fact, even smooth 
potential fields can give rise to dramatic confining effects. 
Now we consider a generic confining potential of the form
\BEQ
A(x) = \frac{x^{2 \alpha} }{2 \alpha}
\EEQ
with $\alpha>0$. The effective potential $B(x)$ reads
\BEA
B(x) &=& \frac{x^{2 \alpha}}{2 \alpha} + \frac{\tau}{2}x^{4 \alpha - 2}  + \\ \nn
&-& \beta^{-1} \ln \left[ 1 +  \tau (2 \alpha - 1) x^{2 \alpha - 2} \right] \; .
\EEA
and the derivative is
\BEA
B^\prime(x) &=& x^{2 \alpha -1 } + \frac{\tau}{2}(4 \alpha - 2) x^{4 \alpha -3} + \\ \nn
&-&\frac{\tau}{\beta} \frac{ (2 \alpha -1) (2 \alpha - 2)x^{2 \alpha -3} }{1 + \tau (2 \alpha -1) x^{2 \alpha - 2}} \ , .
\EEA
\subsubsection{Harmonic trapping \textcolor{black}{of AOUPs}}
The harmonic trapping is recovered for $\alpha=1$.
As it has been proved experimentally and verified in simulations \cite{Maggi14}, harmonic potentials lead to a generalization of the equipartition theorem. 
This is due to the fact that the effective energy takes the simple form
\BEQ
B(x) = \frac{1 + \tau }{2} x^2 - \beta^{-1} \ln (1 + \tau )
\EEQ
that is the energy of an harmonic oscillator where the natural frequency $\omega_0$ frequency results shifted from $\omega_0=1$
to $\omega_0 =  1 + \tau $.  This result means that no accumulation at the boundaries occurs in the case of AOU particles trapped through a harmonic potential.

\subsubsection{Anharmonic trapping  \textcolor{black}{of AOUPs}}
Now we are going to show that the effective equilibrium picture reproduces the accumulations of active particles at the 
boundary of a confining container. 
For sake of simplicity, we consider the case $\alpha=2$, i. e., a soft anharmonic confining potential instead of a container with hard boundaries. 
The effective potential $B(x)$ reads
\BEQ \label{bofx}
B(x) = \frac{x^4}{4} + \frac{\tau}{2} x^{6} - \beta^{-1} \ln \left[ 1 + 3 \tau x^{2}  \right]  \; .
\EEQ
Now a density profile $n(x)$ that is peaked around $x_0=$ turns to be unstable. 
For rationalizing that we compute the 
second derivative of $B(x)$ that is
\BEQ 
B^{\prime \prime}(x) = 3 x^2 + 15 \tau x^4 + \frac{36 x^2 \tau^2}{\beta ( 1 + 3 \tau x^2)^2} - \frac{6 \tau }{\beta ( 1 + 3 \tau x^2)}  \; ,
\EEQ
as one can immediately check, the configuration $x_0=0$ turns to be unstable since $B^{\prime \prime} (0 )= -\frac{6 \tau}{\beta}  < 0$, i. e., as soon as $\tau \neq 0$, $x_0=0$ is not a minimum anymore. 
In terms of the effective potential $B(x)$, the emerging phenomenology can be interpreted as a spontaneous symmetry breaking.
To prove that we consider the small $\tau$ expansion of Eq. (\ref{bofx}) that is
\BEQ \label{landau_1d_acc}
B(x) = -3 \tau x^2 + (\frac{1}{4} + \frac{9 \tau^2}{2}) x^4 + o(x^6) \; .
\EEQ
As one can see, if we think at the potential $A(x)$ in terms of a Landau theory, the potential corresponds 
to a mean-field theory $\frac{a }{2} x^2 + \frac{ x^{4} }{4}  $ at
the critical point, i. e., $a=0$. Now, for $\tau>0$, the system is described by an effective Landau energy that
is given by Eq. (\ref{landau_1d_acc}). If we interpret the coefficient of the $x^2$ term as a mass it turns to
be unphysical since it is negative. This means that the original vacuum of the theory, i. e., $x_0=0$, is not a minimum
anymore and thus $B(x)$ spontaneously breaks the symmetry $x\to-x$ that was satisfied by $A(x)$.
\begin{figure}[!t]
\includegraphics[width=1.\columnwidth]{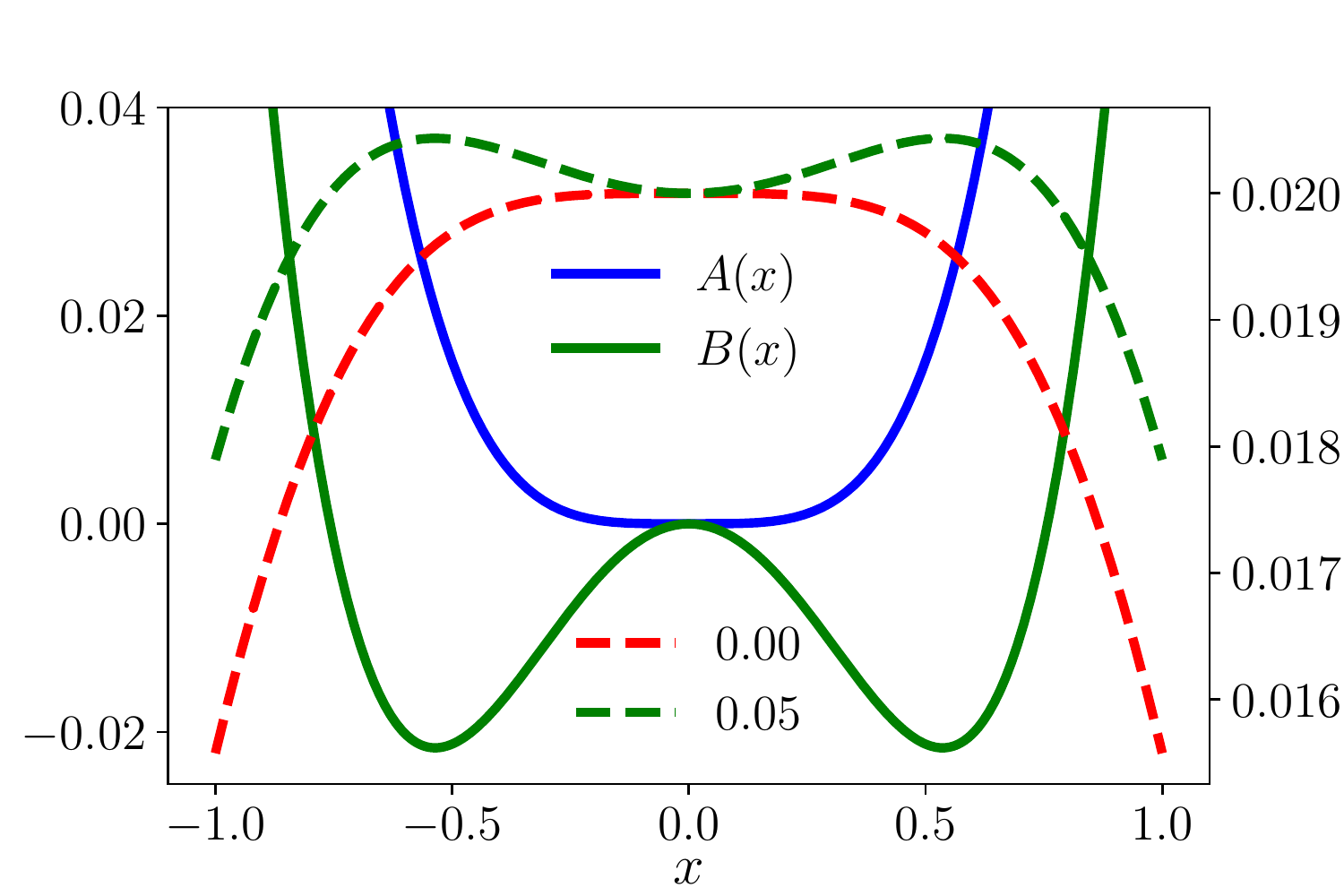} \\
\includegraphics[width=1.\columnwidth]{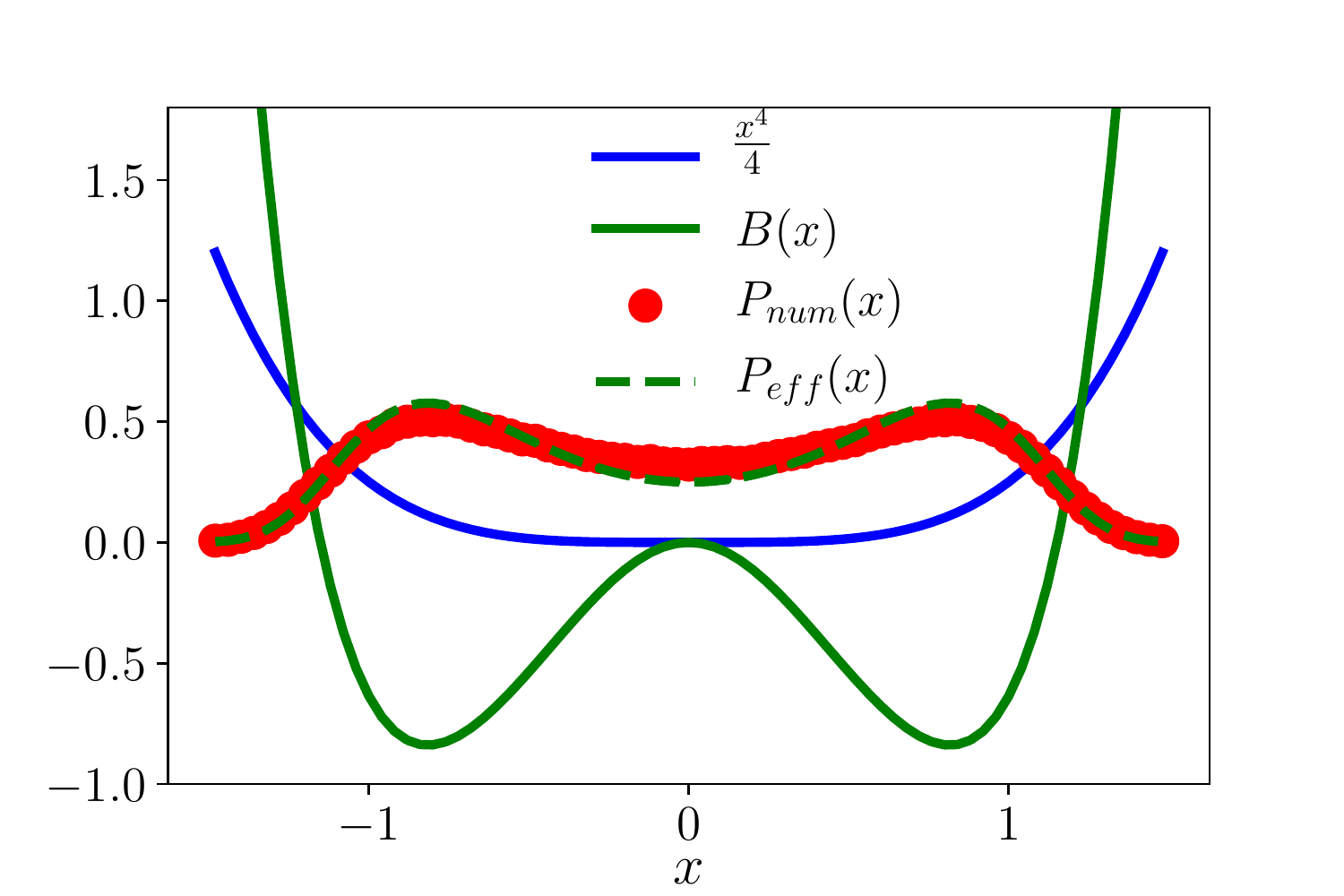}
\caption{ Upper panel:  A gas of active particles embedded into an anharmonic trap in one spatial dimension, $A(x)=\frac{x^4}{4}$ (solid blue curve).
$B(x)$ is the effective potential felt by a particle when $\tau=0.05$. Dashed curves refer to the probability distribution functions.
The effective potential spontaneously breaks the symmetry $x\to-x$.
Lower panel: Comparison between numerical simulations \textcolor{black}{of the actual nonequilibrium dynamics} (red circles) and effective theory (dashed green curve). The solid blue curve is the
anharmonic trapping potential, the solid green curve indicates the effective hamiltonian describing the system. 
The simulation parameters are $\tau=1$, $\mu=1$, and $T=1$.
\label{fig:gas} 
}
\end{figure}
%
%
%
Now the new stable configuration of lower energy is $x_0^{1,2}=\pm \sqrt{ \frac{6 \tau}{ 1 + 18 \tau^2}} \equiv \pm v$.
Density profiles at zero temperature \textcolor{black}{take the form} $n(x) \sim \delta(x - x_0^1) + \delta (x - x_0^2) = \delta(x - v) + \delta(x + v) $. A
 finite but small temperature $T=1/\beta$ introduces
a finite variance in the two distributions that remain peaked around $\pm v$. 
Expanding around the new minima $x = v + \delta$ at the linear order in $\tau$ one has
\BEQ
B (\delta x) = 6 \, \delta x^2 \tau + \frac{\delta x^4}{4} + \sqrt{6} \, \delta x^3 \sqrt{\tau} + o(\tau^2) \; .
\EEQ
As one can see, fluctuations $\delta x$ around the minimum $v$ that spontaneously break the symmetry $x \to -x$ acquire
a mass that is linear in $\tau$.
The emerging phenomenology is shown in the upper panel of Fig. (\ref{fig:gas}) where the solid blue line is the central potential
$A(x)$ with $\alpha=2$, the solid green line is the effective potential $B(x)$ that develops two symmetric minima that break 
the symmetry $x \to -x$. The dashed lines are the distribution function $n(x) \sim \exp{ \left[ -B(x) \right] }$ for $\tau=0$, and $\tau=0.05$, (dashed red, and dashed green, respectively).
As one can appreciate, for $\tau=0$, one has the peak of the distribution at $x_0=0$. For $\tau=0.05$, the distribution become double peaked.

Moreover, the  theoretical picture provides also approximation
schemes for obtaining  quantitative predictions. 
\textcolor{black}{We thus performed numerical simulations of a gas of AOUPs embedded into the potential $A(x)$.
Through the numerical integration of the actual dynamics given by Eqs. (\ref{micro_mo}), we sampled steady-state trajectories for computing numerically $P_{num}(x)$.
The comparison between $P_{num}(x)$ and $P_{eff}(x)$ obtained from UCN
}
is shown in the lower panel of  Fig. (\ref{fig:gas}). The red circles represent the histogram
$P(x)$ computed from numerical data, considering $N=4000$ independent runners.
$P_{eff}(x)$ has been computed considering $P_{eff}(x)=\mathcal{N} e^{-B(x)/T} $ with
$\mathcal{N}$ a normalization constant that guaranties $\int dx \, P_{eff}(x) = 1$. Here, the
computation has been performed considering the expression of $B(x)$ given in Eq. (\ref{bofx}), i.e.,
without performing any small $\tau$ expansion.

\section{Many-body interactions: Mean-Field Theory} \label{many-body-theo}
\textcolor{black}{Now we are going to consider the case of two-body interactions.}
With this aim, we perform 
the coarse-grained computation of Eq. (\ref{ucn_many}) in absence of external fields, i. e., $\phi_1=0$.
For sake of simplicity we indicate the two-body potential $\phi_2(\xx,\yy)=\phi(\xx,\yy)$. 
Moreover, for the computation of the determinant that appears in Eq. (\ref{ucn_many}), we recover to a small $\tau$
expansion.
In term of the
density field $\rho(\xx)$,
using the theoretical framework introduced in Eq. (\ref{setup}), we can write the following mean-field model (details of the computation are provided in appendix \ref{ucn_coarse} )
\begin{widetext}
\BEA \label{many_theory}
f(\beta) &=& -\lim_{N,V \to \infty} \frac{1}{\beta V} \ln Z \\ \nn
Z &=& \int \mathcal{D} \hat{\psi}(\xx) \mathcal{D} \psi(\xx)\, e^{-G[\hat{\psi},\psi ]} \sim  e^{-S_{eff}[ \rho(x)] } \\ \nn
-S_{eff} &\equiv&  \frac{1}{2} \int d\xx d\yy \, \rho(\xx) A(\xx,\yy) \rho(\yy)  + 
\frac{\tau}{2} \int d\xx d\yy d\zz \, \rho(\xx) \rho(\yy) \rho(\zz) B(\xx,\yy,\zz)    + \int d\xx \, \rho(\xx) \left[ 1 + \ln \left( \frac{1 - \rho(\xx)}{\rho(\xx)} \right) \right] \\ \nn
A(\xx,\yy)      &\equiv&  \phi(\xx,\yy) - 2 \tau T \p^2_{\xx , \yy} \phi(\xx,\yy) \\ \nn
B(\xx,\yy,\zz)      &\equiv& \p_\zz \phi(\zz,\xx) \p_\zz \phi(\zz,\yy) \; .
\EEA
\end{widetext}
As one can see, the effective action $S_{eff}$ results from three contributions: (i) a two-body interaction with kernel 
$A(\xx,\yy)$, (ii) a three-body interaction whose kernel is $B(\xx,\yy,\zz)$, and (iii) an entropic term that has combinatorial 
origin \cite{vanKampen,barrat2003basic}. 
It is worth noting that, as we will see in details in the next sections, the two-body interaction, that is responsible 
also for the mass term in the corresponding field theory, can change the sign because of the competition between the original
interaction $\phi$ and quadratic terms originated by \textcolor{black}{the small $\tau$ expansion of $\det \mathbf{M}$.}
\textcolor{black}{When it is different to zero, the three-body interaction developed by the effective equilibrium picture explicitly breaks the symmetry $\rho(\xx) \to -\rho(\xx)$.}


\subsection{Central pair potentials}
Eq. (\ref{many_theory}) defines the thermodynamic of a wide class of scalar active systems. In order to make 
quantitative progresses, we have to define the form of the \textcolor{black}{two-body interaction.}
Let us focus our attention in the case of central pair potentials that take the form
$\phi(\xx_i,\xx_j)= \phi( |\xx_i - \xx_j | )=\phi(r_{ij})$
where we have defined $r_{ij} \equiv |\xx_i - \xx_j | $. 
Indicating with the prime the derivative with respect $r$, 
the diagonal part of the hessian matrix reads
\BEA
H_{ii}^{\alpha \alpha} &=& \sum_{l} \left\{  \frac{\phi^{\prime \prime} (r_{li}) }{r_{li}^2 } (x_l^\alpha - x_i^\alpha )^2 + \right. \\ \nn
&& \left. \phi^\prime (r_{li}) \left[ \frac{1}{r_{li}} -  \frac{(x_l^\alpha - x_i^\alpha )^2}{r_{li}^3 } \right] \right\} \; .
\EEA
In terms of the local density field $\psi(\xx)$ we can thus rewrite $H_2$ as follows
\BEA
H_2[\psi] &=& -\tau T \int d\xx d\yy \, \psi(\xx) f( | \xx - \yy | ) \psi(\yy) \\ \nn
f(r) &\equiv& \phi^{\prime \prime}(r) + \frac{\phi^\prime (r)}{r} \left( d - 1 \right) \; .
\EEA
The contribution of $H_1$ becomes
%
%
\BEA \nn
H_1[\psi] =   \frac{\tau}{2} \int d\xx d\yy d\zz \,  B(\xx-\yy, \xx - \zz) \phi^\prime(r_{xz} ) \psi(\xx) \psi(\yy) \psi(\zz) && \\ 
B(\xx-\yy, \xx - \zz) \equiv \frac{ (\xx - \yy) \cdot (\xx - \zz ) }{r_{xy} r_{xz} } \; . \;\;\;\;\;\;\;\;\;\;\;\;\;\;\;  && 
\EEA

%

At the saddle point we have $\langle \psi(\xx) \rangle = \rho(\xx)$ and thus
we can finally write the effective action for central potentials that is
\BEA \label{lgeff_new}
S_{eff} [\rho] &=& \int d\xx d\yy \,  \rho(\xx) A( | \xx - \yy | ) \rho(\yy) + \\ \nn
&+& \frac{\tau}{2} \int d\xx d\yy d\zz \, B( |\xx - \yy| , | \xx - \zz| ) \rho(\xx) \rho(\yy) \rho(\zz) + \\ \nn
&+& \int d\xx \, \rho(\xx) \left[ 1 + \ln \left( \frac{1 - \rho(\xx)}{\rho(\xx)} \right) \right]
\EEA
where the functions $A(r)$ and $B(r,s)$ now reads
\BEA
A(r) &=& \phi(r) - \tau T f(r) \\ \nn
B(r,s) &=& \frac{\mathbf{r} \cdot \mathbf{s} }{r s} \phi^\prime(r) \phi^\prime(s) \; .
\EEA

\subsection{Motility-Induced Phase Separation}

The action $S_{eff}$ in Eq. (\ref{lgeff_new}) describes the phenomenology of
Motility-Induced Phase Separation (MIPS) \cite{cates2012diffusive}
i. e., the ability of active systems to undergo a spinodal decomposition similar to gas-liquid phase separation that happens
in absence of any attractive force. 

For rationalizing that, let us consider the stability of a homogeneous density profile $\bar{\rho}$. As we will see in details
in the next sections, considering homogeneous solutions, at the saddle point we can write $S_{eff}[\bar{\rho}]=V g(\bar{\rho})$, where
the function $ g(\bar{\rho})$ defines the intensive Gibbs free energy.
The stability of $\bar{\rho}$ that
minimizes $g( \bar{\rho})$ depends on the sign of the second derivative of $g$
with respect $\rho$ computed in $\bar{\rho}$. In particular, in the case of van der Waals 
theory, the negative sign of the coefficient $\rho^2$ ensures the possibility that a homogeneous
density profile may become unstable in a certain region of the phase diagram \cite{vanKampen}. 
If one neglects attractive interactions, the coefficient of the quadratic term turns to be positive
and thus there is not hope to observe a spinodal decomposition.

In the case of active particles, $A(r)$ takes two contributions. The first one is due to the two-body
central potential $\phi(r)$. This is the standard term that one has in the equilibrium theory: it
is attractive or repulsive depending on the type of potential. The second one is given by the function $f(r)$ and is linear in $\tau$, i. e.,  it disappears for $\tau=0$. 
The important thing is that the function $f(r)$ can be always seen as an attractive potential \cite{Farange15}. The intensity of the attraction is tuned by $\tau$.
However, as we will discuss in the next section, in the case of repulsive potentials, there is a threshold value of $\tau$ for observing MIPS.
This fact implies a reentrant phase diagram and thus, at small enough $T=\frac{v^2 \tau}{d}$, 
i. e., at small enough self-propulsion 
velocities, MIPS disappears. 
%


\section{Mean-field theory for Scalar Active Systems} \label{mean-field-theo}
In this section, we will discuss the mean-field solutions of Eq. (\ref{lgeff_new}) assuming homogeneous density profiles. 
In this way, we describe the phase coexistence in active fluids through a van der Waals-like equation. We discuss the general
case of active particles interacting via a central potential. We thus specialize our computation in both cases, repulsive potentials that
give rise to MIPS, and Lennard-Jones potentials.
Since the effective equilibrium theory represented by Eq. (\ref{lgeff_new}) depends on spatial derivatives of the interacting potential $\phi(r)$, 
it is important that $\phi(r)$ is repulsive on short distances but also a smooth function, in order to have well defined first and second derivatives with respect $r$.
\begin{figure}[!t]
\includegraphics[width=1.\columnwidth]{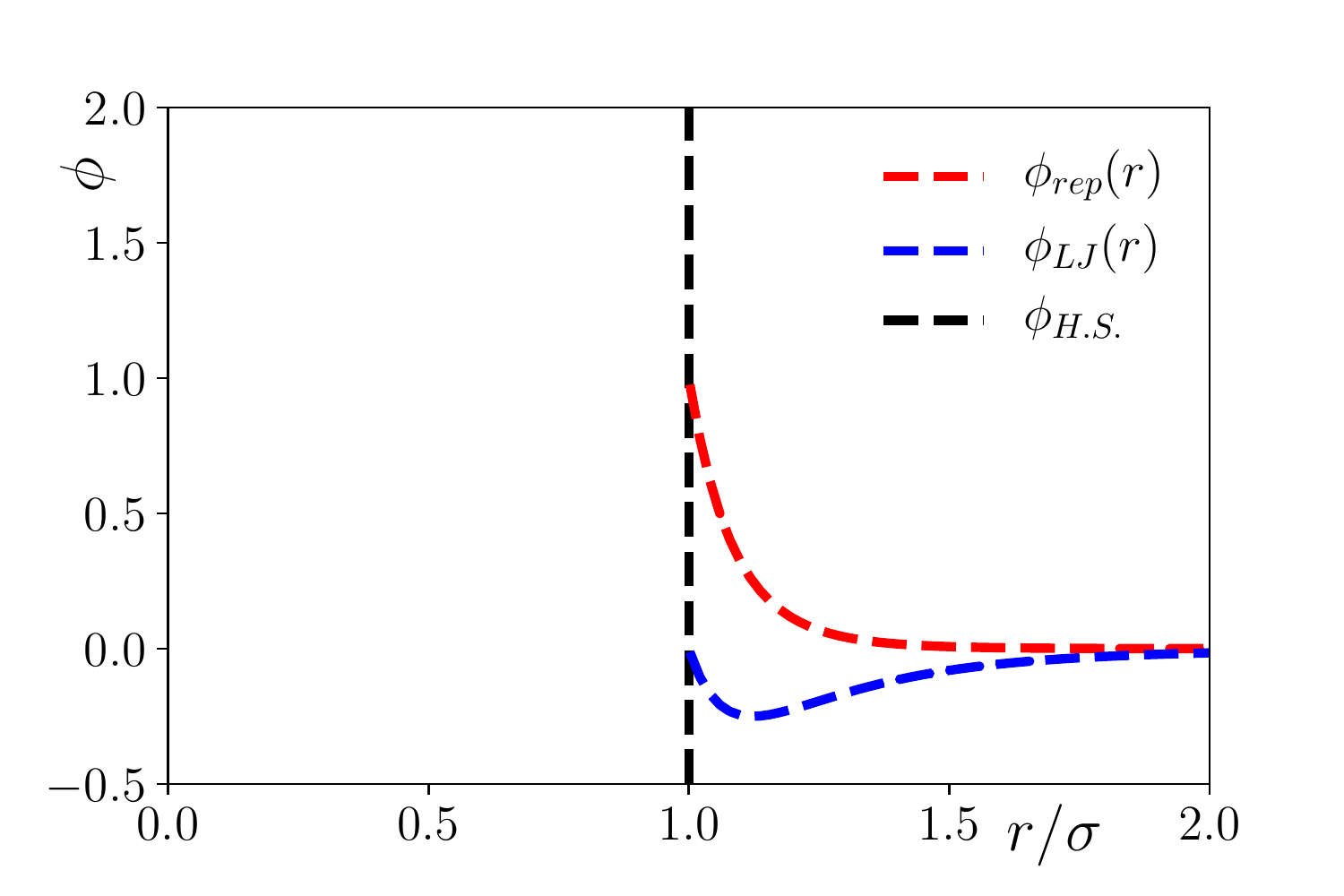}
\caption{ Interaction potentials. We consider an interaction potential $\phi = \phi_{H.S.} + \phi_{smooth}$ that is composed by a hard core repulsion (dashed black line) and a 
smooth interaction $\phi_{smooth}$. For investigating MIPS, the smooth part consists in a purely repulsive potential (dashed red line). We also consider a Lennard-Jones
potential (dashed blue line).  
\label{fig:pot} 
}
\end{figure}

With this aim, we consider the following central potential that results from two contributions, the first one represents a hard-core repulsion that
provides well defined excluded volume effects, the second one is a smooth function that ensures well-defined derivatives. We consider the following function
\BEQ
\phi(r) = \phi_{H.S.}^{\sigma} + \phi_{smooth}(r)
\EEQ
where $\sigma$ is the particle radius. The smooth part $\phi_{smooth}(r)$ is a continuous function of class $\mathcal{C}^{\infty}$, as
it is shown in Fig. (\ref{fig:pot}). We can thus write
\begin{align}
 \phi_{H.S.}^{\sigma} (r) = \left\{
 \begin{array}{cc}
 0      & r > \sigma \\
 \infty & r \leq \sigma \\ 
 \end{array} \right.
\end{align}
and 
\begin{align}
 \phi_{smooth} (r) = \left\{
 \begin{array}{cc}
 0      & r \leq \sigma \\
 \phi(r) & r > \sigma \; .\\ 
 \end{array} \right.
\end{align}
\begin{figure*}[!t]
\includegraphics[width=1.\columnwidth]{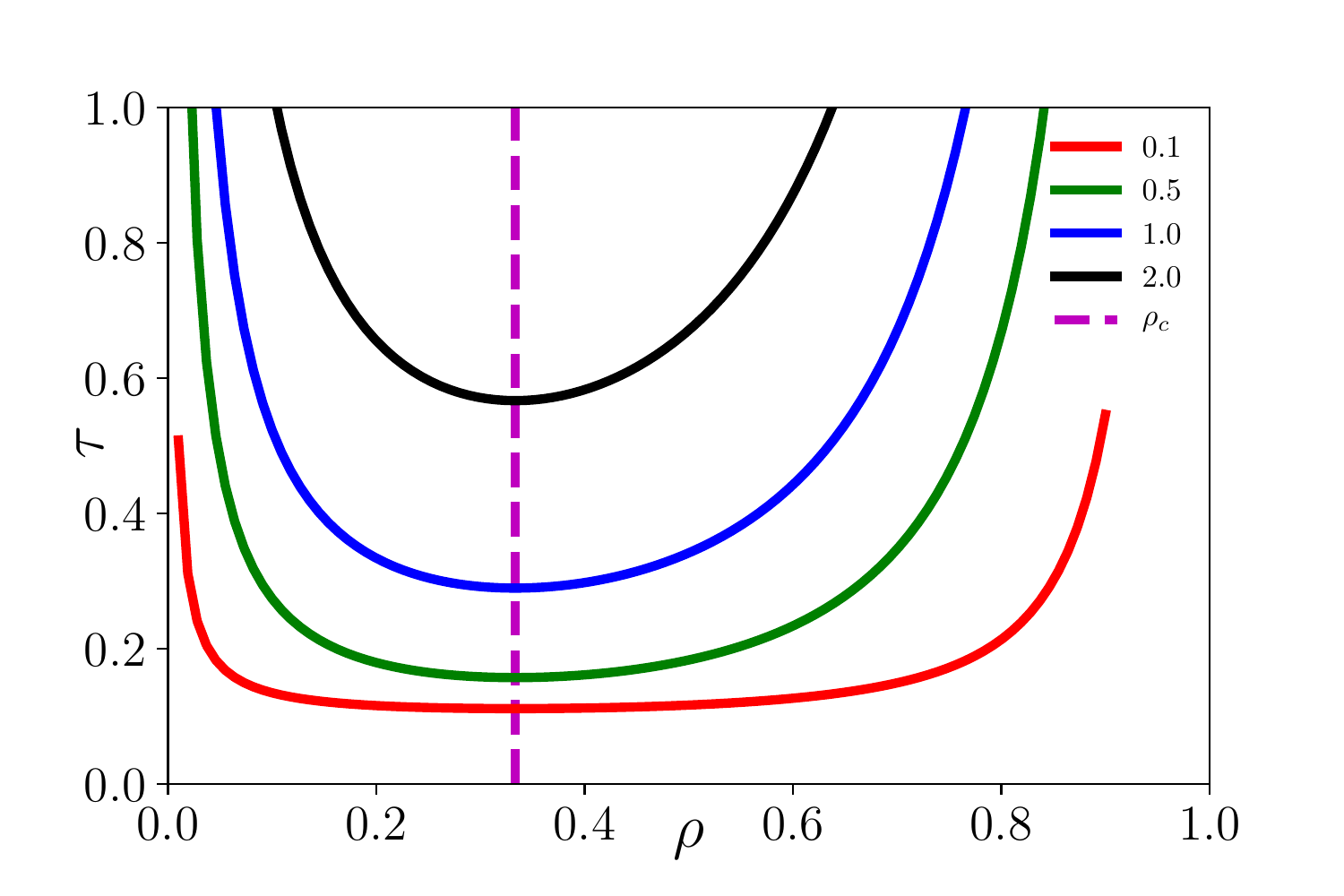}
\includegraphics[width=1.\columnwidth]{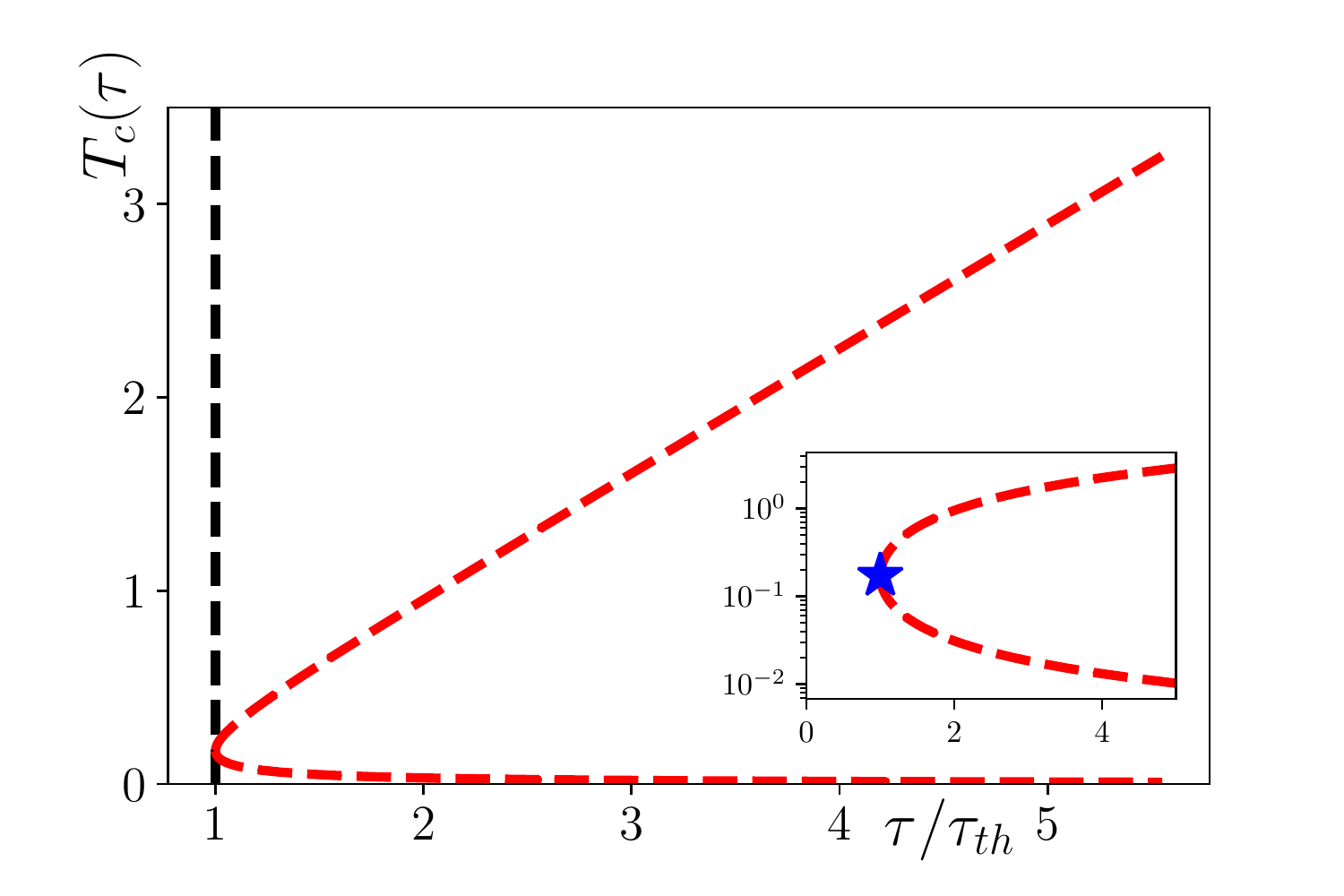}
\caption{  Phase diagram for purely repulsive potentials (MIPS). Left Panel: Phase diagram in the plane $\tau$ vs $\rho$ with $\tau$.
Different lines refer to different noise strength $T$. The dashed magenta line indicates the critical density $\rho_c=1/3$. The solid curves are the spinodals. 
Right Panel: Critical temperature as a function of the persistence time $\tau$ (dashed red line), dashed black vertical line indicates
the threshold value $\tau_{th}$. In the inset we show a zoom of the reentrant region, the star indicates $\tau_{th}$.
\label{fig:pd_rep} 
}
\end{figure*}

Homogeneous density profiles $\rho$ are described by the 
mean-density of the system that is $\rho(\xx) = \rho = N/V$. 
We can
thus rewrite the partition function in Eq. \ref{lgeff_new} as follows
\BEQ
Z \sim e^{-Vg(\rho)} \; ,
\EEQ
where the free energy $g(\rho)$ reads
\BEA \label{free_energy_homogeneous}
-g (\rho) &=& \alpha(\tau,\beta) \rho^2 + \frac{\tau b}{2} \rho^3 + \rho \left[ \ln\frac{1-\rho}{\rho} + 1 \right] \\ \nn
b &\equiv& \int d\xx d\yy d\zz \frac{(\xx - \yy) \cdot (\xx - \zz ) }{r_{xy} r_{xz} } \phi^\prime(r_{xy}) \phi^\prime(r_{xz}) \\ \nn
\alpha(\tau,\beta) &=& \Omega(d) \int dr r^{d-1} \left( \beta \phi(r) -\tau f(r) \right) \\ \nn
f(r) &\equiv& \phi^{\prime \prime}(r) + \frac{ \phi^\prime(r) }{r} (d-1)  \; .
\EEA
In the case of homogeneous density profiles due to spherical particles, 
the coefficient $b=0$ for symmetry reasons. The theory reduces to an effective van der Waals model defined by
the following free energy
\BEQ \label{vdw_new}
-g( \rho ) = \alpha(\tau,\beta) \rho^2  + \rho \left[ \ln\frac{1-\rho}{\rho} + 1 \right] \; .
\EEQ
As one can see, Eq. (\ref{vdw_new}) is precisely the Van der Waals free energy where the effect
of interaction and motility are reabsorbed into the coefficient $\alpha(\tau,\beta)$ \cite{barrat2003basic}.  Homogeneous density profiles
turn to be stable whenever $\alpha(\tau,\beta) \geq 0$. It is worth noting that, in the case of equilibrium systems, the coefficient
is always positive in the case of purely repulsive potentials and thus there is no hope to observe spinodal decompositions at
equilibrium without attractive forces.

\subsection{Repulsive potentials: MIPS critical point and inverse melting}  

For a purely repulsive potential as in the case of $\phi_{smooth}(r) = (\frac{\sigma}{r} )^{12}$, the coefficient $\alpha$ reads
\BEQ \label{alpha_mips}
\alpha(\tau,\beta) = \Omega_d \sigma^d \left[ \frac{\beta}{d - 12} - \frac{12 \tau }{\sigma^2 } \right]  \; , 
\EEQ
with $\Omega_d \equiv \frac{2 \pi^{d/2}}{\Gamma(\frac{d}{2} )}$, and $\Gamma(x)$ the Gamma function.
As we said in the previous section,  $\alpha$ is always positive for $\tau\to0$ indicating that there is not way to observe spinodal decomposition in equilibrium systems. 
For $d<12$, the negative and positive values of $\alpha$ are bounded  by the curve
\BEQ
\beta_{th}(\tau) = \frac{12 (12 - d)}{\sigma^2} \tau \; .
\EEQ
We can write the pressure $P(\rho)$ that is
\BEQ
P(\rho) = \rho \left( \alpha(\beta,\tau) \rho + \frac{1}{\beta \rho (1 - \rho )} \right) \; .
\EEQ
The phase diagram can be obtained considering the solution of $\p_\rho P=0$ that provides the coexistence curve $\tau(T,\rho)$. The computation in
$d$ spatial dimensions and $\sigma=1$ brings to
\BEQ
\tau(T,\rho) = \frac{d \, T^2-2 \rho ^3+4 \rho ^2-2 \rho -12 T^2}{24 (d-12) (\rho -1)^2 \rho  T} \, .
\EEQ
The phase diagram in two spatial dimensions is shown in Fig. (\ref{fig:pd_rep}), left panel. 
The location of the critical point can be obtained considering the solutions of $\p_\rho P=\p_\rho^2 P=0$ that individuate 
the point $\rho_c = \frac{1}{3}$ and $\tau_c(T)=\frac{27 \,d \,T^2-324 T^2-8}{96 (d-12) T}$. As one can see, we obtain the same
critical density of van der Waals theory.

We can also compute the critical temperature as a function of the correlation time of the noise
\BEQ
T_c(\tau) = \frac{16 \tau }{9} \pm \frac{2 \sqrt{2}}{9(d-12)} \sqrt{ (d-12) \left[  3 +  32 (d - 12) \tau^2 \right]} \; .
\EEQ
The behavior of  $T_c(\tau)$ for $d=2$ is shown in the right panel of Fig. (\ref{fig:pd_rep}). It is worth noting that
there is a threshold value $\tau_{th}$ of $\tau$ for having spinodal decomposition. 
Below $\tau_{th}$,
i. e., for $\tau < \tau_{th}$, the system does not undergo a phase transition. Moreover, the critical line $T_c(\tau)$ shows
a reentrance meaning that, decreasing $T$ at fixed $\tau$, the system undergoes an order-to-disorder transition at $T_1=T_c(\tau^1)$
and then, as $T$ decreases below $T_2=T_c(\tau^2) \leq T_1$, the system goes back to a disordered phase passing from liquid to gas. 
The reentrance in the phase diagram is highlighted in the inset of Fig. (\ref{fig:pd_rep}), right panel.


\subsection{Lennard-Jones potentials: Active Gas-Liquid transition}  
\begin{figure*}[!t]
\includegraphics[width=1.\columnwidth]{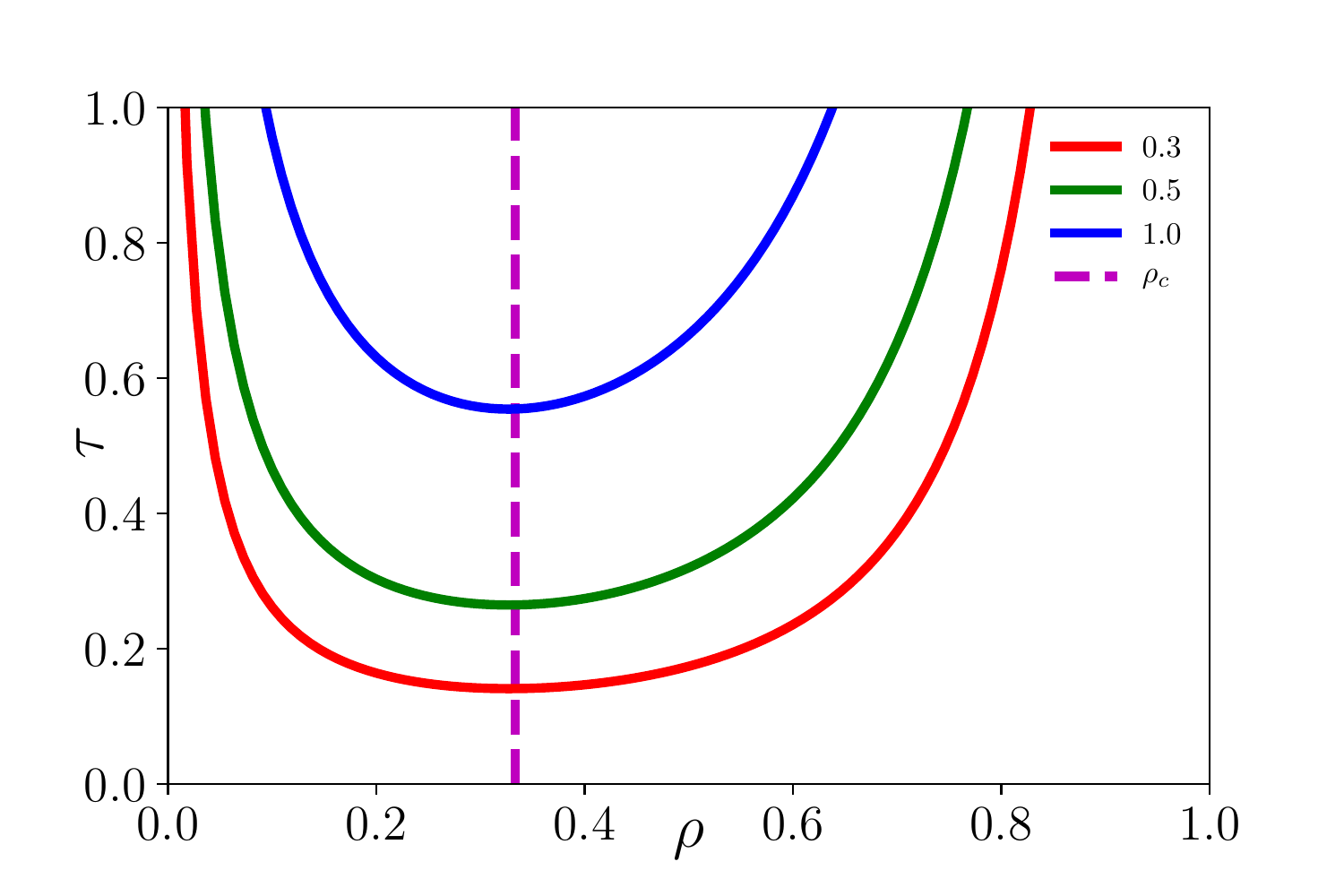}
\includegraphics[width=1.\columnwidth]{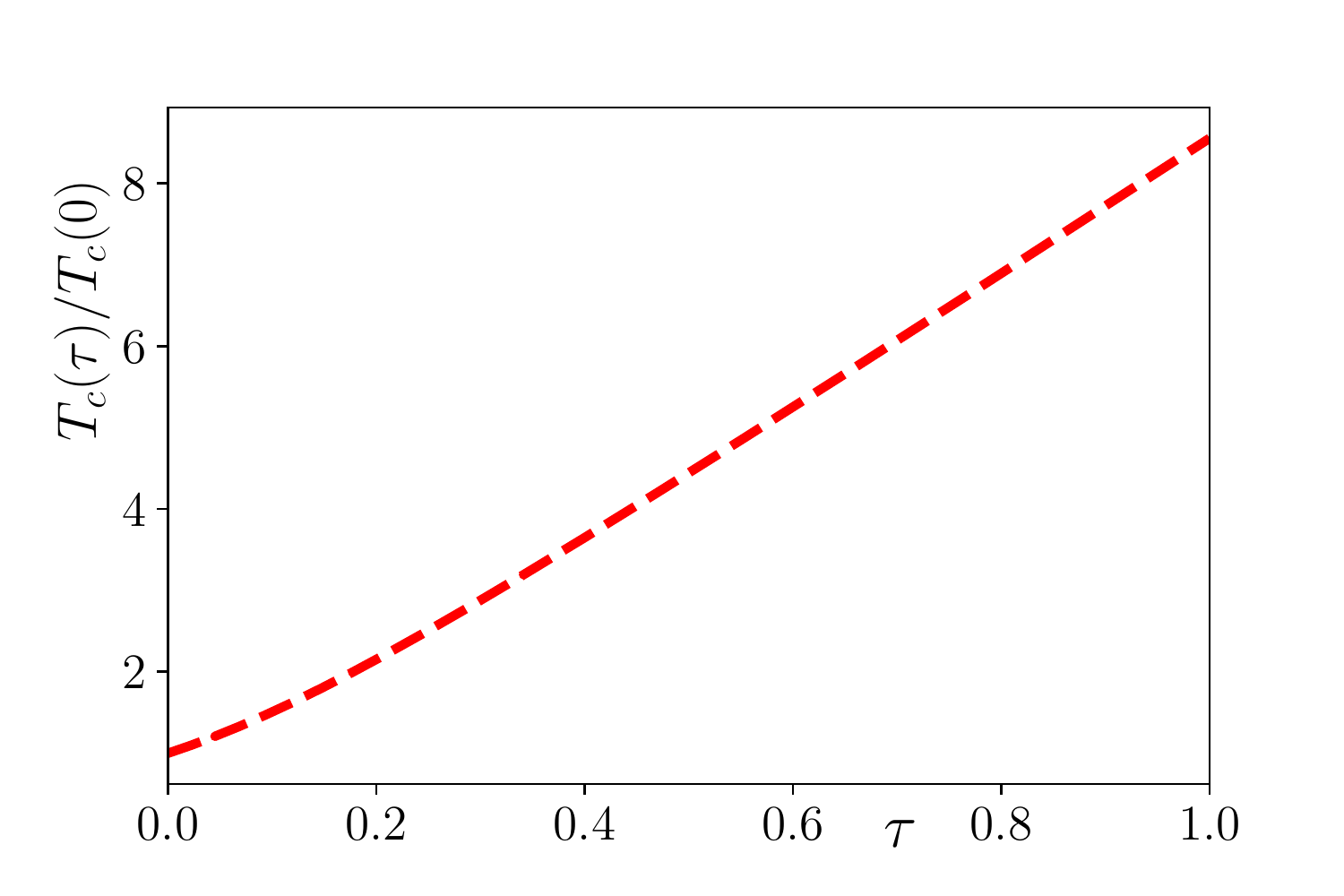}
\caption{  Phase diagram of Lennard-Jones Active Fluids. Left Panel: Phase diagram in the plane $\tau$ vs $\rho$.
Different lines refer to different noise strength $T$. The dashed magenta line indicates the critical density $\rho_c=1/3$. The solid curves are the spinodals. 
Right Panel: Critical temperature as a function of the persistence time $\tau$.
\label{fig:pd_LJ} 
}
\end{figure*}

Now we consider mechanical interactions due to pair potentials that are repulsive at small distances and attractive on a finite range.
Focusing our attention to the case of Lennard-Jones \textcolor{black}{interactions}, one has
\BEQ
\phi_{smooth}(r) = ( \frac{\sigma}{r} )^{12} - (\frac{\sigma}{r})^6 \, .
\EEQ
Again, for $\tau=0$ the system reduces to a Lennard-Jones fluid in equilibrium at temperature $T$.
In equilibrium, one recovers the standard van der Waals theory that locates the critical point at $\rho_c=1/3, \, T_c= 8 \alpha / 27$.
Also in the case of LJ active fluids , for $\tau \neq 0 $, the critical density remains the same while the critical temperature becomes a function of $\tau$, 
i .e., $T_c=T_c(\tau)$. Considering the solutions of $\p_\rho P(\rho_c)=0$, we compute the critical temperature that turns to be
\begin{widetext}
\BEQ
T_c(\tau) = \frac{4 \left(\sqrt{(d-12) (d-6) \left(4 (d-12) (d-6) (d (d+16)-108)^2 \tau ^2+9 (d-14)^2 (d-8)^2\right)}-2 (d-12) (d-6) (d (d+16)-108) \tau \right)}{9 (d-14) (d-12) (d-8) (d-6)} \, .
\EEQ
\end{widetext}
Since LJ fluids undergoes gas-liquid coexistence in equilibrium, one has $T_c(0) \neq 0$.

The resulting phase diagram is shown in Fig. (\ref{fig:pd_LJ}). The solid lines in the left panel represent the spinodal lines, different 
colors refer to different values of $T$. The dashed purple line is the critical density. 
In the right panel it is shown the behavior of the critical temperature as a function of $\tau$, i.e., 
$T_c(\tau)$.

\subsection{Universality class of Motility-Induced Phase Separation}
For investigating the universality class of MIPS within the framework presented here, we have to expand the effective action $S_{eff}$ around
the critical density $\rho_c$ and thus we write $\rho(x) = \rho_c + \varphi(x)$, where $\varphi(x)$ represents a density fluctuation near the transition. 
In doing that, let us introduce the Gibbs free energy $G[\rho] = S_{eff}[\rho]$.
Moreover, we have to introduce the chemical potential $\mu$ that guarantees density fluctuations around the critical point, i.e., we have to consider 
the Legendre transform of $G[\rho] \to G[\rho] - \mu \rho$. In this way, expanding $G$ in power of $\varphi(x)$, one has the cancellation of the linear
term in $\varphi(x)$ that is balanced by the chemical potential term $\mu \varphi(x)$. 
 \textcolor{black}{Around the critical point for $b=0$,}
the expansion $G= \sum_l \frac{1}{l!} a_l \varphi^l$ contains only even power of $\varphi$.
and thus \textcolor{black}{it takes} the form $G[\varphi] = \int d\xx \, \left[ \frac{c}{2} (\nabla \varphi)^2 + \sum_{l=1}^{N_l} \frac{1}{2l!} a_{2l} (\varphi(\xx))^{2l} \right]$.
As a result, the effective theory reduces to a Landau-Ginzburg $\varphi^4$ theory that puts MIPS in the universality class of the Ising model.

However, if we consider elongated particles or density fluctuations that are not isotropic, the term  $b \varphi^3$
might be different from zero \textcolor{black}{ providing an additional source for the breaking of $\varphi \to -\varphi$ symmetry even close to the critical density}.
As a consequence, we might observe deviations from the Ising universality class, \textcolor{black}{in the sense that the critical point might be replaced by a first-order transition}.
This prediction of the mean-field theory has strong consequences in several experimental situations. For instance,
in the case of {\it Myxococcus xanthus} where a MIPS-like phase separation has been observed \cite{AdamPRL}.
In the next section, we will discuss predictions about the theory for $b \neq 0$.
%

\section{MIPS in presence of anisotropic interactions} \label{mips-phi3}
\begin{figure}[!t]
\includegraphics[width=1.\columnwidth]{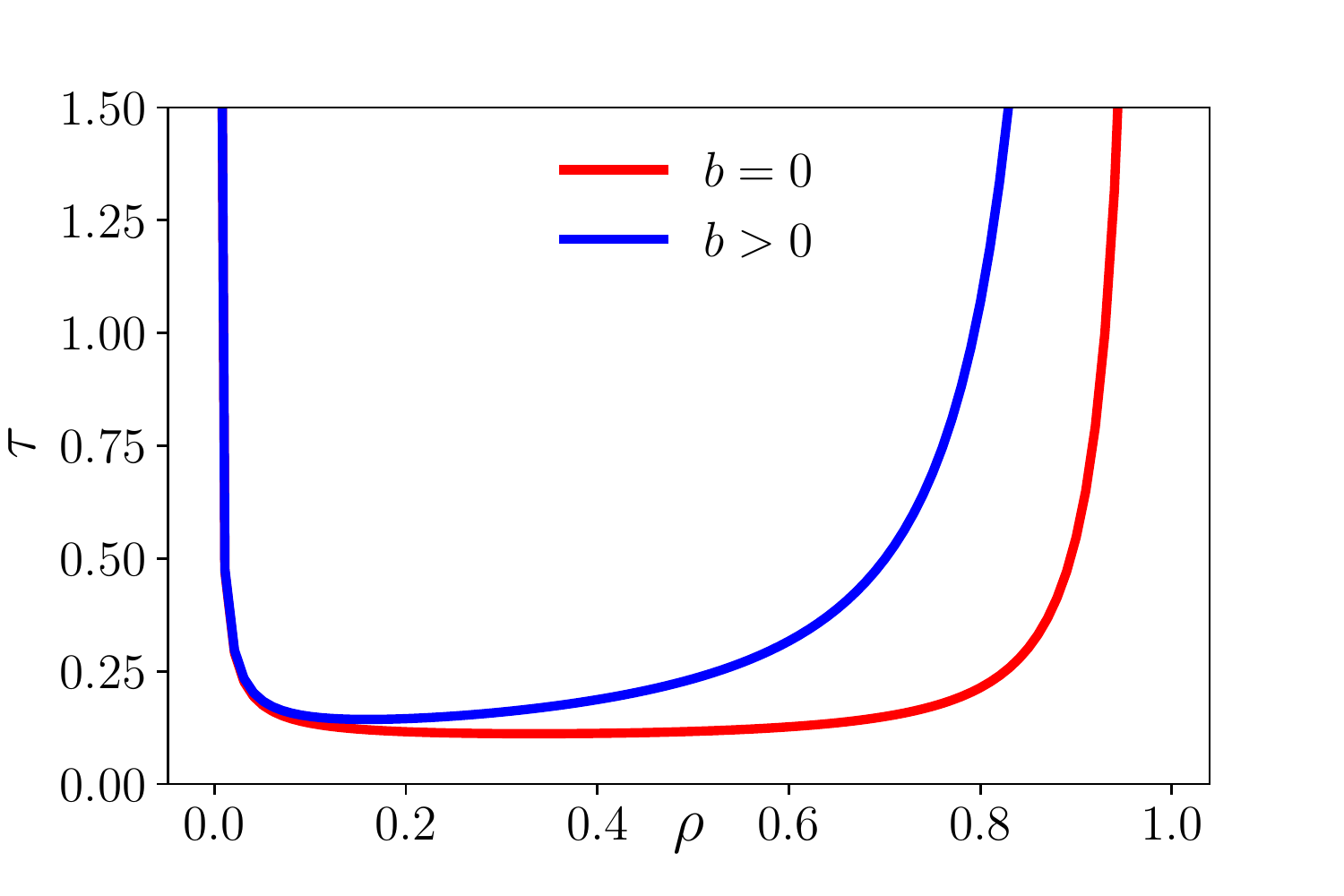} \\
\includegraphics[width=1.\columnwidth]{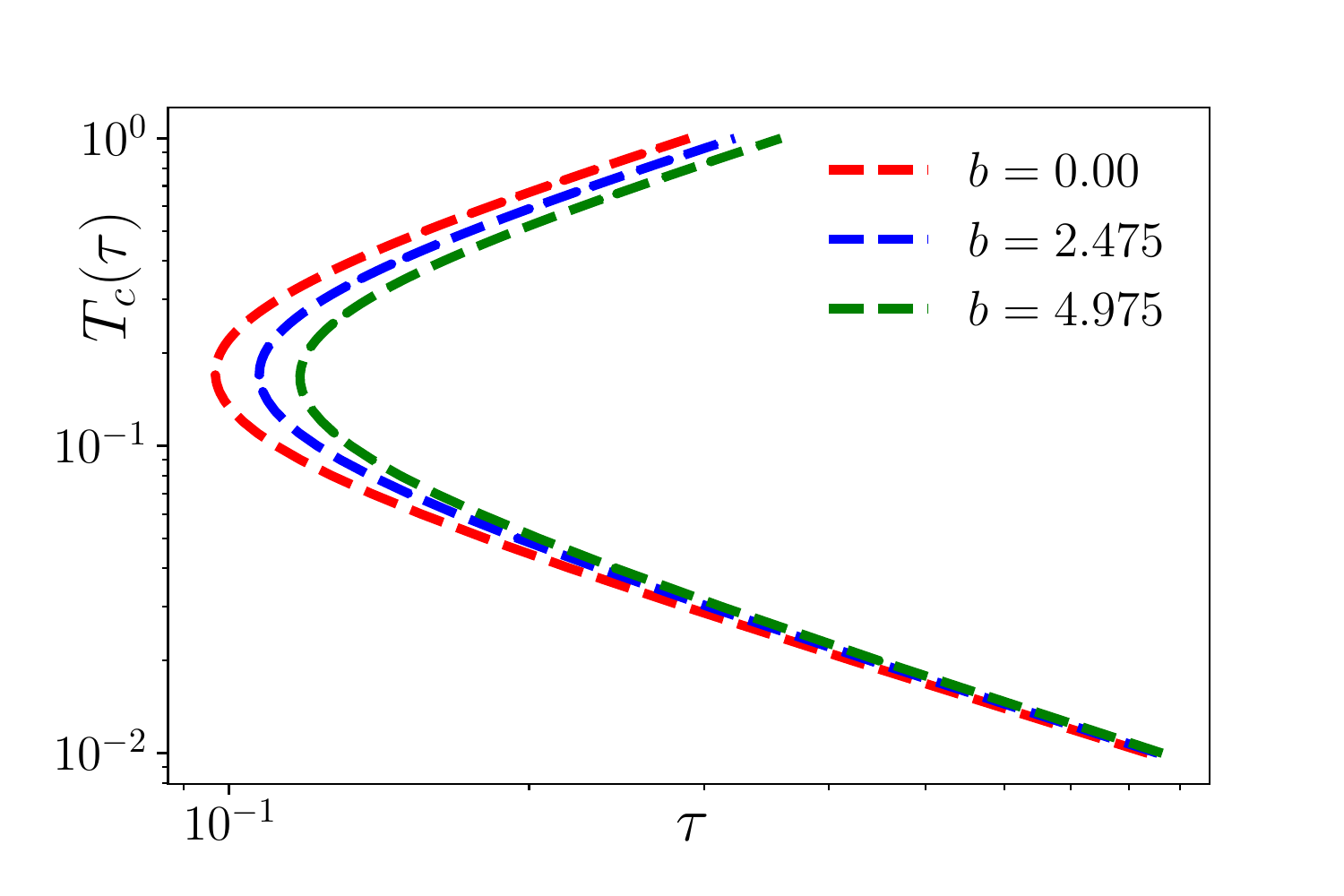}
\caption{ Upper panel. The coexistence curve Eq. (\ref{coe_mips_b}) in presence of asymmetric terms in the Landau-Ginzburg free energy.
Phase diagram in the $\tau$ vs $\rho$ plane with $b=0$ (red curve) and $b=8$ (blue curve). Lower panel. MIPS Critical point $T_c$ as a function of
$\tau$ for as the asymmetry  parameter $b$ increases.
\label{fig:tc_mips_b} 
}
\end{figure}
When we have performed a mean-field approximation on the partition function Eq. (\ref{lgeff_new}), we observed that, in the case
central pair potential, the integral in front of the contribution $\rho^3$ in Eq. (\ref{free_energy_homogeneous}) vanishes identically for
symmetry reasons. It is worth noting that, \textcolor{black}{if the pair potential has not been taken isotropic, i. e., as in the case of elongated particles, }
the integral might assume a finite value.  
In this section, we use again
 Eq. (\ref{free_energy_homogeneous}) as a starting point, i. e., we consider homogeneous solutions of the saddle-point equations. 
 We thus consider a phenomenological coarse-grained theory 
\textcolor{black}{with }
 $b$ 
 as an external 
 control parameter that tunes the degree of anisotropy of the two-body interaction. 
 In this way, we can provide a qualitative estimate of the effect of particle asymmetries on spindally decomposed 
Active Systems. 
We will start by focusing our attention in the case of a  system undergoing MIPS and thus described by $\alpha(\beta,\tau)$ given by Eq. (\ref{alpha_mips}). 
Looking at the solution of $\p_\rho P(\rho)=0$, we can compute the mean-field 
coexistence curves $\tau(\rho,T,b)$ that now depends also on \textcolor{black}{the control parameter} $b$. 
Considering $d=2$, we obtain the following expression for the coexistence curve
\BEQ \label{coe_mips_b}
\tau(\rho,T,b) = \frac{\rho (2 \rho - \rho^2 - 1)- 5 T^2 }{15 T \rho (\rho - 1)^2 (b \rho - 8)}
\EEQ
that reduces to the previous results for $b=0$.
The behavior of Eq. (\ref{coe_mips_b}) is shown in Fig. (\ref{fig:tc_mips_b}), upper panel. As one can appreciate, 
the coexistence region for $b>0$ (blue curve) occupies a smaller area with respect the case of $b=0$. This effect
can be rationalized looking at the free energy Eq. (\ref{free_energy_homogeneous}). The term $b \rho^3$ increases the energy and
thus the system tends to minimize the value of $\rho$. 
\textcolor{black}{This term, in the coexistence region, renormalizes the value of the $\varphi^3$ term that is naturally presented considering the Landau expansion in that region.
However, close to the critical point, where the $\varphi^3$ term of the Landau theory goes to zero, $b \varphi^3$ provides an extra contribution that is in general different from zero.}
In the case of a $\varphi^3$ theory where one has an explicit breaking of $\varphi\to-\varphi$ symmetry, one of the two phases become
energetically preferred. This is signaled by the presence of a metastable minimum in the corresponding Landau-Ginzburg free energy.  
The details are given in the Appendix (\ref{LGPHI3}). It is worth noting that, depending on the value assumed by $b$ with respect to
$\alpha$, the $\varphi^3$ term might destroy the second-order transition that would be
replaced by a first-order one.

\begin{figure}[!t]
\includegraphics[width=1.\columnwidth]{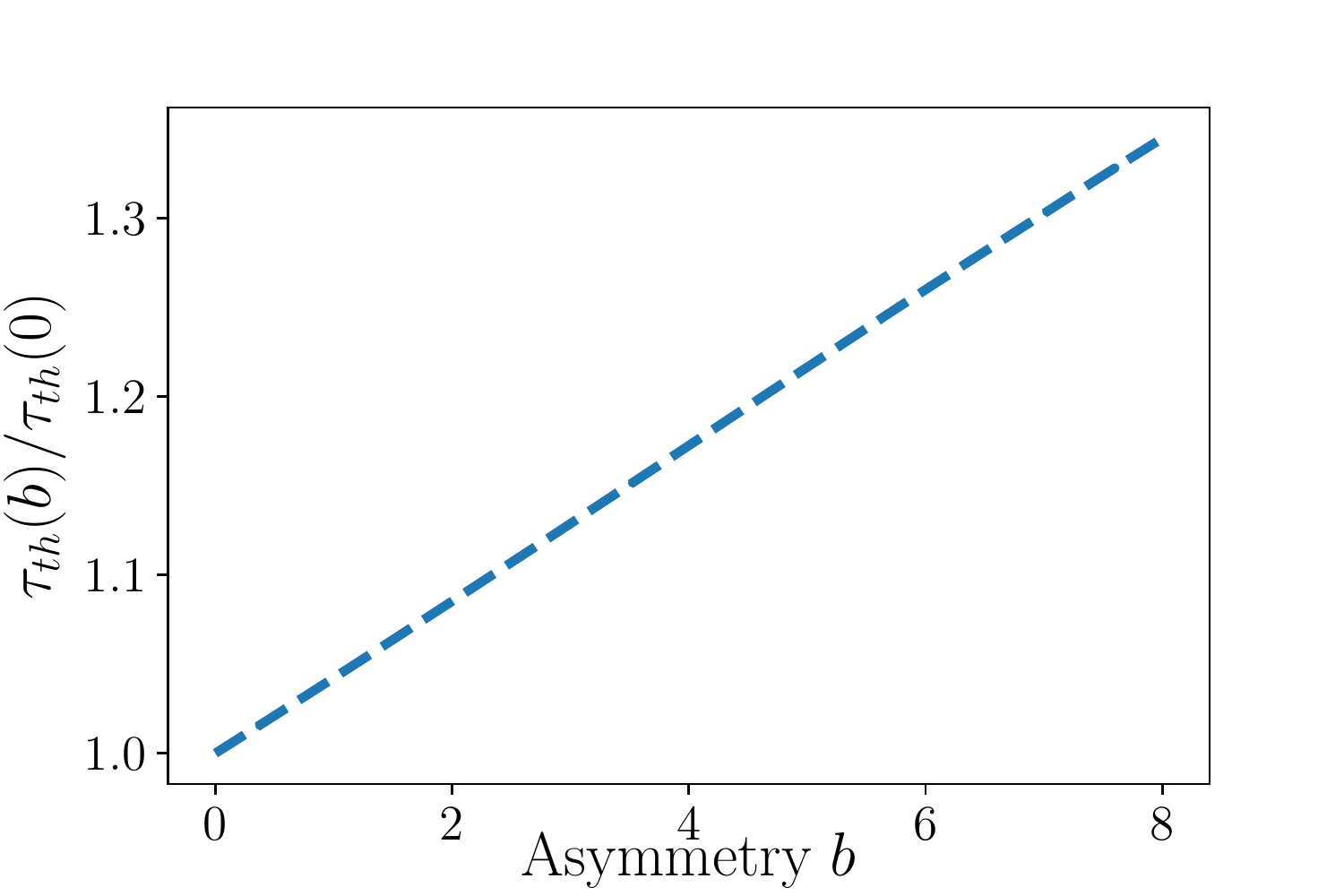}
\caption{ Threshold value $\tau_{th}$, i., e., the minimum value of the persistence time for observing a spinodal decomposition, as a function
of the asymmetry parameter $b$. 
\label{fig:th_mips_b} 
}
\end{figure}

Using Eq. (\ref{coe_mips_b}), we can compute numerically the critical line $T_c(\tau,b)$, i. e., 
the location of the critical point as a function of the 
remaining control parameters. The result is shown in Fig. (\ref{fig:tc_mips_b}), lower panel, for $b=0.00,2.475,4.975$. Again, the critical curve
remains reentrant, however, the asymmetric interaction tends to move the transition at higher $\tau$ values. This effect is made evident 
looking at the shift of the threshold value $\tau_{th}$ as a function of $b$, as it is shown in Fig. (\ref{fig:th_mips_b}).

\section{Discussion and Conclusions} \label{sec-disc}
In this paper,
using the machinery of Statistical Field Theory, we have developed a unified theoretical framework 
 for studying  peculiar non-equilibrium phenomena in scalar Active Matter, i. e., Active System described on large scales by a scalar order parameter $\varphi$. We have derived an effective equilibrium action that contains non-local terms responsible for \textcolor{black}{phase separation} 
 in Active Systems. The non-local terms can be tackled using a mean-field approximation that, in the presented framework,
 turns to be well defined, in the sense that we do not introduce uncontrolled approximation. The mean-field picture is obtained 
 considering the saddle-point solution of the corresponding field theory. Finally, looking at homogeneous solutions and
 the stability of homogeneous configurations, we are able to compute the phase diagram of the system. 
We have focused our attention on three phenomena that concern the condensation of active particles: 
\begin{enumerate}[(i)]
\item Accumulation of active particles at the boundaries of a container.
\item Motility-Induced Phase Separation.
\item Gas-liquid phase transitions in Lennard-Jones Active Fluids. 
\end{enumerate}
It turns out that the phenomenology of scalar Active Matter is captured 
by the effective equilibrium field theory based on UCN approximation. Non-equilibrium phase transitions as MIPS 
are due to non-local many-body contributions in the effective action. 
As a general result, looking the Active System at criticality, scalar Active Matter is described by the following Landau-Ginzburg free energy
\BEQ
f_{LG}^{Active}(\varphi) = \frac{a(T,\tau)}{2} \varphi^2 + \frac{b(\tau)}{3} \varphi^3 + \frac{c}{4} \varphi^4 \; .
\EEQ
 In the case of repulsive potential, the coefficient $a(T,\tau)$ is positive for $\tau=0$, i. e., when the system is in equilibrium. In the case of Lennard-Jones 
systems, $a(T,\tau)$ changes sign at $T_c$ in equilibrium ($\tau=0$). For $\tau>0$, $a(T,\tau)$ vanishes along a critical line $T_c(\tau)$.
The shape of the critical line depends on the microscopic details of the system. In the case of purely repulsive systems, 
$a(T,\tau)=0$ at the MIPS critical point that, at the mean-field level, turns to be in the Ising universality class. 
%
%
According to this picture, in the case $b(\tau)=0$, MIPS results from the spontaneously symmetry breaking
of $\varphi \to -\varphi$ symmetry. We obtained that MIPS takes place above a threshold value of persistence time $\tau_{th}$. Moreover, 
MIPS  is characterized by a reentrant phase diagram in the $T$ vs $\tau$ plane, i. e., above $\tau_{th}$. 
This means that the system evaporates into a gas state by decreasing the driving force.

On the other hand, when $b(\tau)\neq0$, $f_{LG}^{Active}(\varphi)$ predicts also the presence of a
$\varphi^3$ term that explicitly breaks the $\varphi \to -\varphi$ symmetry. We showed that the cubic 
term \textcolor{black}{should} be taken into account when active particles \textcolor{black}{interact through an anisotropic potential.}
e. g., in the case of elongated swimmers. 
\textcolor{black}{As a possible scenario, }
MIPS critical point can be destroyed.
In this case, it might be 
substituted by a first-order phase transition where a metastable MIPS state is nucleated at a higher effective temperature. This state will become 
eventually the stable one at smaller effective temperature. 
Looking at the properties of the system inside the coexistence region, the system is described by a van der Waals-like equation in both
cases, MIPS and Active Lennard-Jones fluid. \textcolor{black}{It is worth noting that the mean field picture developed here predicts MIPS at low persistence times.
This is basically due to the fact that we are considering a small $\tau$ expansion for computing $\det \mathbf{M}$ in Eq. (\ref{ucn_many}).
This is surely in quantitative disagreement with numerical evidences. However, the mean-field picture qualitatively reproduces the main physical mechanisms responsible for phase-separation
in Active Matter. }

In the case of non-interacting active particles, we showed that the accumulation of particles away from the minimum of the trapping potential
can be rationalized in term of a spontaneous symmetry breaking mechanism. In particular, performing a zero temperature 
approximation, the density distribution $\rho(x) \sim \delta(x - x_\alpha)$ turns to be concentrated at $x_\alpha \neq 0$ as soon
as $\tau \neq 0$.

In conclusion, we have shown that effective equilibrium field theories are suitable for gaining insight into \textcolor{black}{condensation phenomena} typical
of Active Systems.

\section*{Acknowledgments} 
MP acknowledges funding from Regione Lazio, Grant Prot. n. 
85-2017-15257 ("Progetti di Gruppi di Ricerca - Legge 13/2008 - art. 4").
This work was also supported by the Joint Laboratory on ``Advanced and Innovative Materials'', ADINMAT, WIS-Sapienza (MP).
We thank Umberto Marini Bettolo Marconi for his critical reading of the manuscript.

 \appendix
 \section{$n-$body interactions in equilibrium Statistical Mechanics} \label{theory}
In this Appendix, we discuss the theoretical framework that we have employed for performing the coarse-graining of the UCN hamiltonian defined in Eq. (\ref{ucn_many}).
Let start with the more general problem of $n-$body interactions in hamiltonian systems, i. e.,  the hamiltonian given in Eq. (\ref{nbody}).
We focus our attention to system described by a scalar field, i. e., the density $\rho(r)$.
As a standard starting point \cite{barrat2003basic}, we consider the identity 
\BEQ 
1 = \int dr \, \delta(r - r_i) \, , \forall i
\EEQ 
and thus we can rewrite the interactions in Eq. (\ref{nbody}) as follows
\begin{widetext}
\BEA 
\sum_{i=1}^N \phi_1(r_i) &=& \sum_{i=1}^N \int dr \, \phi_1(r) \delta(r - r_i)  =  \int dr \, \phi_1(r) \sum_i \delta(r - r_i) \\ \nn 
\sum_{i,j}^{1,N} \phi_2(r_i,r_j) &=& \sum_{i,j} \int drdr^\prime \, \phi_2(r,r^\prime) \delta(r - r_i) \delta(r-r_j) = \int drdr^\prime \, \phi_2(r,r^\prime) \sum_{i,j}\delta(r - r_i) \delta(r-r_j) \\ \nn 
\sum_{i,j,k} \phi_3(r_i,r_j,r_k) &=& \sum_{i,j,k} \int dr dr^\prime dr^{\prime \prime} \phi_3(r,r^\prime,r^{\prime \prime}) \delta(r - r_i) \delta (r - r_j) \delta(r - r_k) =  \\ \nn
&=& \int dr dr^\prime dr^{\prime \prime} \phi_3(r,r^\prime,r^{\prime \prime}) \sum_{i,j,k} \delta(r - r_i) \delta (r - r_j) \delta(r - r_k)
\EEA 
\end{widetext}
It is worth noting that $N = \sum_i \int dr\, \delta(r - r_i) = \int dr \, \sum_i \delta(r - r_i)$. 
The density field $\rho(r)$ is provided by the following relation
\BEQ 
\rho(r) \equiv \langle \delta(r - r_i) \rangle = \frac{\int \prod_i dr_i \, e^{-\beta \mathcal{H} + \beta \mu N} \delta(r - r_i) }{Z}
\EEQ 
In order to perform a coarse-graining of the microscopic model, we introduce the field 
$\psi(r)$ through the identity
\BEQ 
1 = \int \mathcal{D} \psi(r) \delta \left[ \psi(r) - \sum_i \delta(r - r_i) \right] \, ,
\EEQ 
we can thus rewrite the partition function in terms of the coarse-grained scalar field $\psi(r)$ as follows
\begin{widetext}
\BEA 
Z &=& \int \prod_i \frac{dr_i}{\lambda^N} \int \mathcal{D} \psi(r)  \delta \left[ \psi(r) - \sum_i \delta(r - r_i) \right] e^{-S[\psi] - \int dr \, b(r) \psi(r)}\\ \nn 
b(r) &\equiv& \beta \phi_1 (r) - \beta \mu \\ \nn 
S[\psi] &\equiv& \frac{1}{2}\int dr dr^\prime \, \psi(r) \Delta^{-1} (r,r^\prime) \psi(r^\prime) + V[\psi] \\ \nn 
\Delta^{-1} (r,r^\prime) &\equiv& \beta \phi_2 (r,r^\prime ) \\ \nn
V[\psi] &\equiv& \sum_{k \geq 3 } \frac{1}{k !} \int dr_1 ... dr_k \beta \phi_k (r_1,...,r_k) \psi(r_1) ... \psi(r_k) \; .
\EEA 
\end{widetext}
Now, using standard techniques in Statistical Field Theory \cite{qft-ne,ZinnJustin},  we are going to employ the following representation for the delta distribution
\BEQ 
\delta(x) = \int_{s_0 - \ii  \infty}^{s_0 + \ii \infty} \frac{ds}{2 \pi \ii } e^{s x} \, .
\EEQ 
We can thus represent the delta-functional in $Z$ introducing an auxiliary field $\hat{\psi}(r)$ 
and writing
\BEQ 
\delta \left[ \psi(r) - \sum_i \delta(r - r_i)  \right] = \int \frac{\mathcal{D} \hat{\psi} (r)}{2 \pi \ii} \, e^{\int dr  \hat{\psi}(r) \left[ \psi(r) - \sum_i \delta(r - r_i) \right] } \; .
\EEQ 
Now the partition function reads
\BEA 
Z &=& \int \mathcal{D} \psi(r) \mathcal{D} \hat{\psi} (r) e^{-G[\psi,\hat{\psi}] } \\ \nn 
-G &\equiv& -S[\psi] + \int dr \, \left[ \hat{\psi}(r) - b(r) \right] \psi(r)  + N \ln \int \frac{dr}{\lambda} e^{-\hat{\psi} (r)} \,.
\EEA
The field $\hat{\psi}(r)$ can be shifted in a way  that $\hat{\psi} - b \to \hat{\psi}$ and thus
the partition function becomes
\BEA 
Z &=& \int \mathcal{D} \psi(r) \mathcal{D} \hat{\psi} (r) e^{-G[\psi,\hat{\psi}] } \\ \nn 
-G &\equiv& -S[\psi] + \int dr \, \hat{\psi}(r) \psi(r)  + N \ln \int \frac{dr}{\lambda} e^{-\hat{\psi} (r) - b(r)} \, .
\EEA
For making further progresses, we introduce the following generating functional $W[\hat{\psi}]$
that is
\BEQ \label{new_gen_app}
e^{W[\hat{\psi}]} \equiv \mathcal{N}^{-1} \int \mathcal{D} \psi e^{-S[\psi] + \int dr \, \hat{\psi}(r) \psi(r)} 
\EEQ
where we have introduced the normalization $\mathcal{N}\equiv \int \mathcal{D} \psi\, e^{-S[\psi]}$. As one can appreciate, the auxiliary field $\hat{\psi}$ in Eq. (\ref{new_gen_app}) plays the role of external source in a quantum field theory. It is worth noting that $W[\hat{\psi}] = O(N)$. We can now introduce $W[\hat{\psi}]$ in the expression of $Z$ obtaining 
\BEQ \label{zeta_func_app}
Z = \mathcal{N} \int \mathcal{D} \hat{\psi}  \, e^{W[\hat{\psi}] + N \ln \int \frac{dr}{\lambda} e^{\hat{\psi}(r) - b(r)}} \; .
\EEQ 
Finally, we define the functional $F[\hat{\psi}]$ through the relation
\BEQ
F[\hat{\psi}] = W[\hat{\psi}] + N \ln \int \frac{dr}{\lambda} \, e^{-\hat{\psi}(r) + b(r)} \;. 
\EEQ
\section{Fluctuations} \label{fluctua}
In this section, we compute the fluctuations around the saddle-point configuration $\psi_{SP}$.
Let $\hat{\psi}_{SP}(r)$
be the field configuration that makes the action $F[\hat{\psi}]$ stationary, we can write
\BEQ 
\hat{\psi} (r) = \hat{\psi}_{SP}(r) + \Delta \hat{\psi} (r) \; .
\EEQ 
With the field configuration $\hat{\psi}_{SP}(r)$ the solution of the saddle-point equation
\BEQ 
\left. \frac{\delta F }{ \delta \hat{\psi}(r)} \right|_{\hat{\psi}=\hat{\psi}_{SP}} = 0 \; .
\EEQ 
We can now expand the effective action $F[\hat{\psi}]$ up to the second order obtaining
\BEQ 
F[\hat{\psi}] = F[\hat{\psi}_{SP}] + \frac{1}{2} \int dr dr^\prime \, \Delta \hat{\psi} (r) \mathcal{G}(r,r^\prime) \Delta \hat{\psi} (r^\prime)
\EEQ 
where the quadratic part is defined as follows
\BEQ \label{quad}
\mathcal{G}(r,s) \equiv \left. \frac{\delta^2 F[\hat{\psi}]}{\delta \hat{\psi}(r) \delta \hat{\psi}(s)} \right|_{SP} \, .
\EEQ 
It is worth noting that the computation of the stationary configurations brings to the following
self-consistency equation
\BEQ \label{stationary}
\frac{\delta}{\delta \hat{\psi}_{SP}} W[\hat{\psi}_{SP}] = N \frac{e^{-\hat{\psi}_{SP} - b(r) }  }{\int dr \, e^{-\hat{\psi}_{SP}  - b(r) }  } \,.
\EEQ 
Now we compute the Gaussian fluctuations around the stationary solution.
In order to perform the computation we notice that
\BEQ 
\frac{\delta F}{\delta \hat{\psi}(r)} = \frac{\delta W}{\delta \hat{\psi}(r)} + N \frac{1}{z} \frac{\delta z}{\delta \hat{\psi}(r)}
\EEQ 
where we have defined 
\BEQ 
z \equiv \int \frac{dr}{\lambda} e^{- \hat{\psi}(r) + b(r)} \, .
\EEQ 
The functional derivatives of $z$ with respect the field $\hat{\psi}(r)$ gives the following relations
\BEA 
-\frac{\delta z}{\delta \hat{\psi}(r)} &=& e^{- \hat{\psi} (r) + b(r)} \\ \nn
\frac{\delta^2 z}{\delta \hat{\psi}(r) \delta \hat{\psi}(s)} &=& e^{- \hat{\psi} (r) + b(r)} \delta(r - s)
\EEA

Now we define $\rho(r)$ and $G(r,s)$ as follows
\BEA 
\left. \frac{\delta W}{\delta \hat{\psi} (r)} \right|_{SP} &\equiv& \rho(r) \\ \nn
\left. \frac{\delta^2 W}{\delta \hat{\psi}(r) \delta \hat{\psi}(s) } \right|_{SP} &\equiv& G (r,s)
\EEA 
and thus the quadratic part in Eq. (\ref{quad}) can be rewritten as 
\BEQ \label{fluct}
\frac{\delta F^2[\hat{\psi}]}{\delta \hat{\psi}(r) \delta \hat{\psi}(s)} = G (r,s) - \rho(r) \delta(r-s) + \frac{1}{N} \rho(r) \rho(s) \; .
\EEQ 
After introducing $G(r,s)$ and $\rho(r)$, we can finally write the partition function in terms of two contributions: the first one is due to 
the saddle-point configuration $\hat{\psi}_{SP}$ of the field, the second one due to fluctuation around it. We can thus write 
\BEA
Z &=& Z_{SP} Z_{fluct} \\ \nn
Z_{SP} &\equiv&  \mathcal{N} e^{W[\hat{\psi}_{SP}] + N \ln z[\hat{\psi}_{SP}]} \\ \nn
Z_{fluct} &\equiv& \int \mathcal{D} \hat{\psi}\, e^{\frac{1}{2} \int drds\, \hat{\psi}(r) \left[ G (r,s) - \rho(r) \delta(r-s) + \frac{1}{N} \rho(r) \rho(s) \right] \hat{\psi}(s)}
\EEA
where we have changed notation $\Delta \psi \to \hat{\psi}$.
Through the propagator  $G^{-1}(r,s)$ we can perform the following transformation
\BEQ 
\hat{\psi}(r) = \ii \int ds \, G^{-1}(r,s) \varphi(s)
\EEQ 
Employing such a change of variable one obtains
\BEA
&&\int dr ds \, \hat{\psi}(r) G(r,s) \hat{\psi}(s) = \\ \nn
&&= - \int dr ds \, \int ds^\prime \, G^{-1}(r,s^\prime) \varphi(s^\prime)
G(r,s) \times \\ \nn 
&&\int ds^{\prime\prime} \, G^{-1}(s,s^{\prime\prime}) \varphi(s^{\prime\prime}) = \\ \nn
&=& \int ds^\prime ds^{\prime\prime} \varphi(s^\prime) G^{-1}(s^\prime,s^{\prime\prime}) \varphi(s^{\prime\prime}) \, .
\EEA  
The second term in Eq. (\ref{fluct}) becomes
\BEA 
&&\int dr \, \hat{\psi}(r) \rho(r) \hat{\psi} (r) = \\ \nn
&=& \int dr ds ds^\prime \, \varphi(s)  G^{-1}(r,s) G^{-1}(r,s^\prime) \rho(r) \varphi(s^\prime) \, ,
\EEA 
and the third term reads
\BEA 
&&\int drdr^\prime \, \hat{\psi}(r) \rho(r) \rho(r^\prime) \hat{\psi}(r^\prime) = \\ \nn 
&=& \int ds ds^\prime \varphi(s) \left[ \int dr \, G^{-1}(r,s) \rho(r) \right] \times \\ \nn &&\left[ \int dr \, G^{-1}(r,s^\prime) \rho(r) \right] \varphi(s^\prime)\, .
\EEA
It is worth noting that the last term in Eq. (\ref{fluct}) gives subd-leading contribution since it
is multiplied by $1/N$. Finally, we have to take into account the Jacobian determinant 
of the transformation $\hat{\psi} \to \varphi $. Putting all together we arrive to
the following expression for the partition function 
\begin{widetext}
\BEA  \label{finale}
Z &=& \frac{\mathcal{N}}{\det G} e^{F[\hat{\psi}_{SP}]}\int \mathcal{D} \varphi \, e^{-\frac{1}{2} \int dr dr^\prime \varphi \mathcal{G} \varphi} \\ \nn 
\mathcal{G} &\equiv& G^{-1}(r,r^\prime) - \int ds \, G^{-1}(s,r) \rho(s) G^{-1}(s,r^\prime) + O(1/N)\\ \nn 
F[\hat{\psi}_{SP}] &\equiv& W[\hat{\psi}_{SP}] + N \ln z[\hat{\psi}_{SP}] \\ \nn
&&\hat{\psi}_{SP} \; : \; \left. \frac{\delta F }{\delta \hat{\psi}} \right|_{SP} = 0 \\ \nn
\rho &=& \frac{\delta }{\delta \hat{\psi}_{SP}} W[\hat{\psi}_{SP}]\, , \;\; G \equiv  \frac{\delta^2 }{\delta \hat{\psi}^2_{SP}} W[\hat{\psi}_{SP}] \\ \nn 
e^{W[\hat{\psi}]} &=& \mathcal{N}^{-1} \int \mathcal{D} \psi \, e^{-S[\psi] + \int dx \, \hat{\psi} \psi} \\ \nn 
\mathcal{N} &\equiv& \int \mathcal{D} \psi \, e^{-S[\psi]} \\ \nn 
S[\psi] &\equiv& \beta H[\psi] - \left. \psi \frac{\delta}{ \delta \psi} \beta H \right|_{\psi=0}
\EEA  
\end{widetext}
In writing Eqs. (\ref{finale}) we have neglected $O(1/N)$ terms that are
\BEQ
\frac{1}{N} \left[ \int ds \, \rho(s) G^{-1} (s,r)\right] \, \left[ \int ds \, \rho(s) G^{-1}(s,r^\prime)\right] \propto \frac{1}{N} \; .
\EEQ 

 \section{Gas in thermal equilibrium in external potentials} \label{1dgas}
Now we consider a gas composed by non-interacting particles in one spatial dimension that are 
embedded into an external potential $A(x)$. The external potential
is a smooth and continuous function of class $\mathcal{C}^{\infty}$. The hamiltonian $H$ reads
\BEQ
H[x_i]=\sum_{i=1}^N A(x_i) \; .
\EEQ 
The partition function is
\BEQ
Z_\beta = \int \prod_i dx_i \, e^{-\beta H[x_i]} \, ,
\EEQ
in terms of the number density field $\psi(x)$ and the auxiliary field $\hat{\psi}(x)$ we can write
\BEA
Z_\beta &=& \int \mathcal{D} \psi(x) \mathcal{D} \hat{\psi} (x) \, e^{-G[\psi,\hat{\psi}]} \\ \nn
G[\psi,\hat{\psi}] &\equiv& \hat{\psi} \cdot \psi + \beta \psi \cdot A - N \ln z[\hat{\psi}] \\ \nn
z[\hat{\psi}] &\equiv& \int dx \, e^{\hat{\psi}(x)} \\ \nn
f \cdot g &\equiv& \int dx \, f(x) g(x) \, .
\EEA
Indicating with $n(x) \equiv \psi(x)_{SP} $ and $\hat{n}(x) \equiv \hat{\psi}(x)_{SP} $ the filed configurations that satisfy the self-consistency equations, we have
\BEQ
n(x) - N \frac{e^{\hat{n} (x)}}{z} = 0 \, ,
\EEQ
and we can thus write
\BEQ
\hat{n}(x) = \ln z[\hat{n}] + \ln n(x) - \ln N \, .
\EEQ 
Since $\int dx \, n(x) = N$ we can finally write
\BEQ
G[n] = \int dx \, n(x) \left[ \beta A(x) + \ln n(x) \right]
\EEQ
and the corresponding self-consistency equation is
\BEQ
n(x) = c e^{-\beta A(x)}
\EEQ
with $c=const.>0$. At small temperature $\beta \to \infty$ the density profile is dominated 
by the minima of $A(x)$ and thus $n(x) \sim \sum_\alpha \delta(x - x_0^\alpha)$ with $x_0^\alpha$ given by
\BEQ
\left. \frac{d A}{dx} \right|_{x=x_0^\alpha}=0 \; , \left. \frac{d^2 A(x)}{dx^2} \right|_{x=x_0^\alpha} \geq 0 \; .
\EEQ

\subsubsection{Soft confining potentials}
As a case study we are going to consider a gas in equilibrium confined through a potential
\BEQ
A(x) = \frac{x^{2 \alpha} }{2 \alpha}
\EEQ
the corresponding density profile is
\BEQ
n(x) = c \, e^{-\beta \frac{x^{2 \alpha} }{2 \alpha} } \; .
\EEQ
At high temperatures $\beta \to 0$ and then $n(x)=c$. The constant is fixed by the normalization, considering the
particles free to move in a segment of length $L$ centered around the origin, one has
\BEQ
\int_{-L/2}^{L/2} dx \, n(x) = c \, L = 1 
\EEQ
and thus $c=\frac{1}{L}$.
At zero temperature $n(x) \sim \delta(x - x_0)$ with $x_0$ given by
\BEQ
\left. \frac{d A}{dx} \right|_{x=x_0}= x_0^{2\alpha -1}  = 0
\EEQ
that has $x_0=0$ as unique and stable solution $\forall \, \alpha > 0$.

\section{Mean-Field Theory of Two-body Interactions}
Here we start with a specific problem that is the computation of the partition function $Z_\beta$ in the case
of a particle system interacting through a pairwise potential.  The hamiltonian reads
\BEQ
H(\pv,\rr)=\sum_{i=1}^N\frac{\pv_i^2}{2 m} + \sum_{i<j} \phi( \rr_i , \rr_j )
\EEQ
and the partition function $Z_\beta$ is
\BEA
Z_\beta&=&\frac{1}{h^{3N} N!} \int \prod_i d\pv_i d\rr_i e^{-\beta H(\pv,\rr)} =  \\ \nn
 &=& \left(\frac{2 \pi m}{\beta}\right)^{\frac{3N}{2}} \frac{1}{h^{3N} N!} X_\beta
\EEA
where we have defined the configurational integral
\BEQ
X_\beta \equiv \int \prod_i d\rr_i e^{-\beta \sum_{i<j} \phi(  \rr_i , \rr_j )} \, .
\EEQ
In order to perform a coarse-graining of the microscopic dynamics, 
we introduce the density field $\rho(\rr)$ that is
\BEQ
\psi(\rr) =  \sum_{i=1}^N \delta (\rr - \rr_i)
\EEQ
we can then rewrite the two-body interaction $\Phi\equiv \sum_{i<j} \phi(  \rr_i , \rr_j )$ in the following way
\BEA
\Phi &\equiv& \frac{1}{2}\sum_{i,j} \phi(  \rr_i , \rr_j   ) = \\ \nn
&=& \frac{1}{2} \int d\rr d\rr^\prime \, \psi(\rr) \phi(  \rr , \rr^\prime ) \psi(\rr^\prime) - N \phi(0)\, .
\EEA
The density field can be forced into the partition function using a delta functional
\BEQ
\int \mathcal{D}\psi(\rr) \, \delta \left[ \psi(\rr) \!-\! \sum_i \delta(\rr - \rr_i) \right] = 1
\EEQ
that brings to the following expression for the configurational integral
\BEA
&&X_\beta = \int \prod_i d\rr_i \mathcal{D} \psi(\rr) \, \delta \left[ \psi(\rr) - \sum_i \delta(\rr - \rr_i) \right] \times \\ \nn 
&& \exp{\left(  -\frac{\beta}{2} \int d\rr d\rr^\prime \psi(\rr) \phi( \rr , \rr^\prime ) \psi(\rr^\prime) \right) } \; .
\EEA
The delta functional can be expressed using an auxiliary field $\hat{\psi}(\rr)$
\BEA
 &&\delta \left[ \psi (\rr) - \sum_i \delta(\rr - \rr_i) \right] = \\ \nn 
 &&=\int \mathcal{D} \hat{\psi}(\rr) \, e^{-  \int d\rr \psi(\rr) \hat{\psi} (\rr) +  \int d\rr \hat{\psi}(\rr) \sum_i \delta (\rr - \rr_i) }
\EEA
and the configurational integral becomes
\BEA \label{xx}
X_\beta &=& 
\int \prod_i d\rr_i \mathcal{D} \psi(\rr)  \mathcal{D}\hat{\psi}(\rr)  \, \exp \left( -  \int d\rr \psi(\rr) \hat{ \psi } (\rr) \right. \\ \nn 
 + &&\left. \int d\rr \hat{ \psi }(\rr) \sum_i \delta (\rr - \rr_i) -\frac{ \beta }{2} \int d\rr d\rr^\prime \psi(\rr) \phi( \rr , \rr^\prime ) \psi(\rr^\prime) \right)  \; .
\EEA
Let us introduce the one-body partition function $z[\hat{ \psi }]$ that is
\BEQ
z[\hat{ \psi }] \equiv \left( \int  d\rr \, e^{\hat{ \psi }(\rr)} \right)^N
\EEQ
and then we can write
\BEA \label{saddle}
X_\beta &=& \int  \mathcal{D} \psi(\rr)  \mathcal{D} \hat{\psi}(\rr) e^{-G[ \psi, \hat{ \psi }]} \\ \nn
G[\psi,\hat{\psi}] &\equiv& \int d\rr \, \psi(\rr) \hat{ \psi} (\rr) -  \ln z[\hat{\psi}] + \\ \nn 
&& + \frac{ \beta}{2} \int d\rr d\rr^\prime \psi(\rr) \phi(  \rr , \rr^\prime ) \psi(\rr^\prime) \; . 
\EEA
In the thermodynamic limit we assume that fluctuations are negligible. In this situations, the integral is dominated by the extreme values of $G$. Performing 
a saddle point approximation, one has
\BEA \label{speqs}
\left. \frac{\delta G}{\delta \hat{ \psi }(\rr)} \right|_{\hat{\psi} = \hat{\rho}\,,\,\psi = \rho} &=& \rho (\rr) - \frac{N e^{\hat{ \rho }(\rr)}}{\int d\rr \, e^{\hat{ \rho }(\rr)} } = 0 \\ \nn
\left. \frac{\delta G}{\delta \psi(\rr)} \right|_{\hat{\psi} = \hat{\rho}\,,\,\psi = \rho}           &=& \hat{ \rho }(\rr) +  \beta \int d \rr^\prime \phi(  \rr , \rr^\prime  ) \rho(\rr^\prime) =0 \; . 
\EEA
Using the first equation, we can write
\BEQ 
\ln \rho(\rr) + \ln z - \ln N - \hat{\rho}(\rr) = 0 \; ,
\EEQ
and thus
\BEQ \label{enne}
 \hat{\rho}(\rr) = \ln \rho(\rr) + \ln z - \ln N \; .
\EEQ
We can rewrite the free energy at the saddle-point as a functional of the density field $\rho(\rr)$.
Now we plug Eq. (\ref{enne}) into Eq. (\ref{saddle}) for obtaining 
\BEA
G[\rho] &=& \int d\rr \, \rho(\rr) \left[ \ln \rho(\rr) + \ln z - \ln N \right] - N \ln z + \\ \nn 
&& + \frac{ \beta}{2} \int d\rr d\rr^\prime \rho(\rr) \phi(  \rr , \rr^\prime ) \rho(\rr^\prime) 
\EEA

and finally 
\BEA \label{gofrho}
&&G[\rho] = \frac{ \beta }{2} \int d\rr d\rr^\prime \, \rho(\rr) \phi( \rr , \rr^\prime ) \rho(\rr^\prime)  + \\ \nn
&& \int d\rr \, \rho(\rr) \ln \rho(\rr) - N \ln N 
\EEA
where we have used the constraint $\int d \rr \rho( \rr )= N$.

In order to study the stability of a solution $(\rho(\rr),\hat{\rho}(\rr) )$, one ha to compute
the hessian matrix $\mathbf{H}$ with components $H_{\alpha,\beta} = \frac{\delta^2 G}{\delta \rho_\alpha (\rr) \delta \rho_\beta (\rr^\prime) } $, where the greek indices takes the values $\alpha=1,2$ with $\rho_1 \equiv \rho$ and $\rho_2  \equiv \hat{\rho}$.
the stability condition is
\BEQ \label{stability}
\left. \det \frac{\delta^2 G}{\delta\rho_\alpha  (\rr) \delta \rho_\beta (\rr^\prime) }  \right|_{(\rho(\rr),\hat{\rho}(\rr) )} \geq 0 \; . 
\EEQ
In the case of a two-body potential one has to consider the eigenvalues of the matrix
\BEQ 
M(\xx,\yy) = \beta \phi(\xx,\yy) + \frac{\delta(\xx - \yy)}{\rho(\xx)} \; .
\EEQ 
The two-body potential $\phi(\xx,\yy)$ is translationally invariant, we can thus write $\phi(\xx,\yy)=\phi(\xx - \yy)$ and the matrix becomes
\BEQ 
M(\xx,\yy) = M(\xx - \yy) = \beta \phi(\xx - \yy) + \frac{\delta(\xx - \yy)}{\rho(\xx)} \; .
\EEQ 
Since the continuous matrix $M(\xx - \yy)$ is
translational-invariant, it is diagonal in Fourier space. Now we study the stability
of a homogeneous density profile $\bar{\rho}$. Using the expressions
\BEA 
\phi(\xx - \yy)    &=& \sum_{\kk} e^{i \kk \cdot (\xx - \yy) } \hat{\phi}_\kk \\ \nn 
\delta(\xx - \yy)  &=& \sum_{\kk} e^{i \kk \cdot (\xx - \yy) } \; ,
\EEA 
and the sums run over wave vectors $\kk = \frac{2 \pi}{L} \mathbf{n}$.
We obtain that homogeneous configurations are stable if
\BEQ 
\beta \hat{\phi}_\kk + \frac{1}{\bar{\rho}} \geq 0 \; .
\EEQ 
Since we are interested on the large scale behavior of the system, we look at $\kk \to 0$ and thus
\BEQ 
\hat{\phi}_0 = \Omega(d) \int dx \, x^{d-1} \, \phi(x), 
\EEQ 
where $\Omega(d)$ results form the integration on the solid angle in $d$ spatial dimensions.
As we will see in the next section, for a Van der Waals gas the integral is negative and thus there is a critical
density above that the homogeneous solution is not stable anymore.

\subsection{Stability of homogeneous density profiles}
Differently from the van der Waals theory, here we did not consider explicitly excluded volume effects. As a consequence, we cannot
observe a spinodal decomposition between a liquid and a gas phase. However, we can still study the  stability of homogeneous
density profiles. 
For homogeneous solutions $\rho(\rr)=Const.=\rho$, assuming translational invariant interactions, we can write
\BEA \label{vdweq}
G[\rho] &=& V g(\rho) \\ \nn
g(\rho) &\equiv& \frac{\beta \alpha}{2} \rho^2 + \rho \ln \rho \\ \nn
\alpha &\equiv& \Omega(d) \int dr \, r^{d-1} \phi(r) \, .
\EEA
The pressure $P(\rho)$ can be computed as
\BEQ
P(\rho) = \rho \frac{\p g}{\p \rho} - g(\rho)
\EEQ
and thus we have
\BEA
P(\rho) &=& \rho \left( \beta \alpha \rho  +1 + \ln \rho \right) -\frac{\beta \alpha}{2} \rho^2 - \rho \ln \rho= \\ \nn
&=& \rho \left( \frac{\beta \alpha}{2} \rho + 1 \right)
\EEA
If we require the thermodynamical stability of the solution $\rho$, we have to compute the second derivative of $g(\rho)$
that is
\BEQ
\frac{\p^2 g}{\p \rho^2} = \beta \alpha + \frac{1}{\rho} \, .
\EEQ
For purely repulsive potentials, $\alpha > 0$ and thus $\frac{\p^2 g}{\p \rho^2} \geq 0$, $\forall \rho \in [0,1]$, i.e., homogeneous density
profiles are always stable in equilibrium systems when attracting forces between particles are negligible.

Considering an equilibrium system where particles interact through a potential that is attractive on short distances, the integral
in the third of Eqs. (\ref{vdweq}) turns to assume negative values,
i. e., $\alpha < 0$. Let us write $\alpha = - |\alpha |$, now the stability of homogeneous profiles
is related to
\BEQ \label{critical_point}
\frac{\p^2 g}{\p \rho^2} = - \beta | \alpha| + \frac{1}{\rho} \, .
\EEQ
that changes sign at $\rho = \rho_c = \frac{1}{\beta |\alpha| } $. The location of the critical point can computed considering the equations
\BEQ
\frac{\p P}{\p \rho} = \frac{\p^2 P}{\p \rho^2}=0
\EEQ
that are
\BEA \label{cri}
&& 1- \beta |\alpha| \rho = 0 \\ \nn
&& -\alpha \beta = 0
\EEA
The system of equations (\ref{cri}) does not have solutions and thus, in the model we have considered, we do not have any critical point.

\subsection{Including excluded volume effects in the mean-field theory} \label{coarse}
For obtaining the van der Waals theory it results convenient to do study the system discretized on a lattice. We follow a standard procedure
that can be found in Refs. \cite{vanKampen,hughes2014introduction,barrat2003basic}. We illustrate the method considering an equilibrium  system composed by
$N$ particles in a volume $V$ in $d$ spatial dimensions. We perform a coarse-graining dividing the systme into a lattice that defines a set of occupation number ${N_i}$ that must satisfy 
the constraint $\sum_i N_i=N$. Each cell occupies a volume $\Delta$ and thus $V = N \Delta$, each particle occupies a volume $\delta$.
We consider the potential $\phi(\rr_i,\rr_j)$ composed by two parts  
\BEQ
\phi(\rr_i,\rr_j) = \phi_{HS}^{\delta} + \phi_{smooth}(\rr_i,\rr_j)
\EEQ
where $\phi_{HS}^{\delta}$ indicates a hard core potential,  i. e.,  each particle is a hard spheres that occupies a volume $\delta$.
The second part $\phi_{smooth}(r)$ is a smooth function of $r$. The hard core potential causes a contraction of the phase
space $\omega(N_i)$ that now is $\omega(N_i)=(\Delta - N_i \delta)^{N_i}$. As we have done before when we have defined 
the local density filed $\psi(\rr)$, through the occupation numbers $N_i$ the energy can be written as
\BEQ
\Phi = \frac{1}{2}\sum_{i,j} \phi_{i,j} N_i N_j \; ,
\EEQ
with $\phi_{i,j} \equiv \phi_{smooth}(\rr_i,\rr_j)$
The configurational integral $X_N$ becomes
\BEA
X_\beta &=& \sum_{\{ N_i\} | \sum_i N_i = N} e^{-G[N_i]} \\ \nn
-G[N_i] &\equiv& \sum_i \left[ N_i \ln \left( \Delta - N_i \delta \right)  - N_i \ln N_i + N_i \right]  \\ \nn
&+& \frac{\beta}{2} \sum_{ij} \phi_{i,j} N_i N_j \; .
\EEA
At the saddle-point one has to find the solution of the set of equations
\BEQ
\ln \left( \frac{\Delta - N_i \delta}{N_i} \right) - \left( \frac{N_i \delta }{\Delta - N_i \delta} \right) + \beta \sum_{i,j} \phi_{i,j} N_j = \mu
\EEQ
where the Lagrangian multiplier $\mu$ guaranties $\sum_i N_i = N$. Considering the uniform solution of the saddle-point equations
that has the form $N_i = \rho \Delta$, $\forall i$ with $\rho = N/V$. Setting $\delta=1$, the free energy becomes
\BEA \label{vdw}
-G[\rho] &=& -V g(\rho) \\ \nn
-g(\rho) &\equiv& \rho \ln\left( \frac{1 - \rho}{\rho} \right) + \rho + \frac{\beta \alpha}{2} \rho^2 \\ \nn
\alpha &\equiv& \sum_{i,j} \phi_{i,j} \Delta \; .
\EEA
Focusing our attention on repulsive potential, we can write $\alpha = - |\alpha|$ and compute the van der Waals equation of state that is
\BEQ
\frac{\beta P(\rho)}{\rho} =  \frac{1}{ \rho -1 } - \frac{\beta \alpha \rho}{2} \;.
\EEQ
The critical point is determined through Eqs. (\ref{critical_point}) that brings to $\beta_c= \frac{27}{4 \alpha} $ and $\rho_c = \frac{1}{3}$.

In order to make in contact Eq. (\ref{vdw}) with Eq. (\ref{gofrho}) we perform the continuous limit $\Delta \to 0$ that brings to
\BEA \label{gofrho}
-G[\rho] &=& \int d\rr \, \rho(\rr) \left[ \ln\frac{1 - \rho(\rr)}{\rho(\rr)} + 1 \right] + \\ \nn
&+& \frac{\beta}{2} \int d\rr d\rr^\prime \, \phi_{smooth}(\rr,\rr^\prime) \rho(\rr) \rho(\rr^\prime) \; .
\EEA
As one can appreciate, Eq. (\ref{gofrho}) has the form of a Density Functional Theory (DFT). For obtaining 
the kinetic term $\nabla \rho$, it is convenient to go back to the discretized action and look at interaction term
\BEQ
\sum_{i,j} \phi_{ij} N_i N_j \; ,
\EEQ
using the identity
\BEQ
N_i N_j = \frac{1}{2} \left( N_i ^2+ N_j^2 \right) - \frac{1}{2} \left( N_i - N_j \right)^2
\EEQ
the interaction term can be written as
\BEQ
\sum_{i \neq j}\phi_{ij} N_i N_j =  \sum_{i\neq j} \phi_{ij} N_i^2 - \frac{1}{2} \sum_{i \neq j } \phi_{ij} (N_i - N_j)^2 \; .
\EEQ
 Let $\rho_i$ be the density in the box $i$ that is $\rho_i = \frac{N_i}{\Delta} $.  We can thus rewrite
\BEA
&&\sum_{i\neq j} \phi_{ij} N_i^2 - \frac{1}{2} \sum_{i \neq j } \phi_{ij} (N_i - N_j)^2  = \\ \nn
&&\sum_{i\neq j} \Delta \Delta \, \phi_{ij} \rho_i^2 - \frac{1}{2} \sum_{i \neq j } \Delta \Delta  \phi_{ij} (\rho_i - \rho_j)^2 \; .
\EEA
 In the continuum limit, i.e., $\Delta \to 0$, we get
 \BEA
&&\sum_{i\neq j} \Delta \Delta \, \phi_{ij} \rho_i^2 - \frac{1}{2} \sum_{i \neq j } \Delta \Delta  \phi_{ij} (\rho_i - \rho_j)^2 \\ \nn
&& \longrightarrow \epsilon \int d\rr  \, \rho(\rr)^2 +  \frac{c}{2} \int d\rr \, ( \nabla \rho )^2 \; .
 \EEA
 The first term in the last equation is nothing else than a mass term in a field theory. 

\section{UCN as an effective equilibrium theory} \label{ucn_coarse}
In this appendix we show the computation that brings to Eq. (\ref{many_theory}).
For sake of simplicity let us change the notation from $\rr_i$ to $\xx_i$ for indicating a
configuration of the system. Now we consider the following effective hamiltonian
\BEA
H_{UCN}[\xx_i] &=& H_0[\xx_i] + H_1[\xx_i] + H_2[\xx_i] \\ \nn
H_0[\xx_i] &=& \frac{1}{2} \sum_{i,j} \phi(\xx_i, \xx_j) \\ \nn
H_1[\xx_i] &\equiv& \frac{\tau}{2} \sum_i \left( \nabla_{\xx_i} H_0 \right)^2 \\ \nn
H_2[\xx_i] &\equiv& - T \ln \det M \\ \nn
M &\equiv& \delta_{ij} + \tau \frac{\p^2 H_0}{\p \xx_i \p \xx_j} \; .
\EEA
The thermodynamics is given as usual
\BEA
F &=& -\frac{1}{\beta} \ln Z_\beta \\ \nn
Z_\beta &=& \int \prod_i d\xx_i \, e^{-\beta H_{UCN}[\xx_i]}
\EEA
Now we introduce the density field $\psi(\xx)$ that is
\BEQ
\psi(\xx) = \sum_i \delta( \xx_i - \xx)
\EEQ 
and the auxiliary field $\hat{\psi}(\xx)$. We can thus write
the thermodynamics as
\BEQ 
Z_\beta = \int \mathcal{D} \psi(\xx) \mathcal{D} \hat{\psi}(\xx) \, e^{-G[\psi, \hat{\psi}]}
\EEQ 
In terms of the field $\psi(\xx)$, the three terms $H_0$, $H_1$, and
$H_2$ can be written as follows.

\paragraph*{$H_0$ term.---}
This is the usual two-body term that can be coarse-grained as follows
\BEQ
H_0[\psi] = \frac{1}{2} \int d\xx d\yy \, \rho(\xx) \phi(\xx,\yy) \rho(\yy)
\EEQ

\paragraph*{$H_1$ term.---}
This term brings to non-linear and non-local interactions in the effective equilibrium theory. In particular, it
is responsible for a $\psi^3$ term that explicitly breaks the symmetry $\psi \to -\psi$. This term can be rewritten as follows
\BEA
H_1[\psi] &=& \frac{\tau}{8} \sum_i \left[ \frac{\p}{\p \xx_i} \sum_{k,l}  \phi(\xx_k, \xx_l) \right]^2 = \\ \nn
&=& \frac{\tau}{8} \sum_i 4 \left[  \sum_{k,l} \frac{ \p \phi(\xx_k, \xx_l)}{\p \xx_k} \frac{\p \xx_k}{\p \xx_i}   \right]^2 = \\ \nn
&=& \frac{\tau}{2} \sum_i \left[  \sum_{k,l}  \frac{ \p \phi(\xx_k, \xx_l)}{\p \xx_k} \delta_{ki}  \right]^2 = \\ \nn
&=&\frac{\tau}{2} \sum_i \left[  \sum_{l} \frac{ \p \phi(\xx_i, \xx_l)}{\p \xx_i}  \right]^2 = \\ \nn
&=& \frac{\tau}{2} \sum_i \left[ \int d\xx \, \psi(\xx) \p_{\xx_i} \phi(\xx_i, \xx) \right]^2 = \\ \nn
&=& \frac{\tau}{2} \sum_i \left[ \int d\xx d\yy \, \psi(\xx) \psi(\yy)  \p_{\xx_i} \phi(\xx_i, \xx)  \p_{\xx_i} \phi(\xx_i, \yy) \right]  = \\ \nn
&=&\frac{\tau}{2} \int d\xx d\yy d\zz \, \psi(\xx) \psi(\yy) \psi(\zz)  \p_{\zz} \phi(\zz ,\xx)  \p_{\zz} \phi(\zz, \yy)  \, .
\EEA
Where we have introduced the notation $\p_\zz \phi(\zz,\xx)\equiv \frac{\p \phi(\zz,\xx)}{\p \zz} $.

\paragraph*{$H_2$ term.---}
In order to perform the coarse-graining of the interactions due to $H_2$, we have to deal with the determinant of a $dN \times dN$ matrix.
For doing that, tet us assume a small $\tau$ limit, in this case, indicating with $\mathbf{H}$ the hessian matrix and
with $\mathbf{1}$ the identity matrix, one has
\BEQ
\det (\mathbf{1} + \tau \mathbf{H}) \simeq 1 + \tau \, Tr \, \mathbf{H}
\EEQ
In this limit we can write
\BEQ
H_2[\psi] = -  \tau T  \int d\xx d\yy \, \psi(\xx) \p_{\xx,\yy}^2 \phi(\xx , \yy) \psi(\yy)
\EEQ
where we have defined $\p_{\xx,\yy}^2 \phi(\xx , \yy)\equiv \frac{\p^2 \phi(\xx,\yy) }{\p \xx \p \yy} $.

For proving that, we write the determinant of a generic matrix $\mathbf{M}$ of elements $M_{ij}$ as
\BEQ 
\det M_{ij} = \int \prod_{i} d\theta_i d\bar{\theta}_{i} \; , e^{\theta_i M_{ij} \bar{\theta}_j}
\EEQ 
where we have introduced two sets of conjugated Grassmann variables $\theta_i$ and $\bar{\theta}_i$ \cite{ZinnJustin}, respectively, that
satisfy the anti-commutation rules
\BEA 
\{ \theta_i, \theta_j \}             &=& 0 \\ \nn 
\{ \bar{\theta}_i, \bar{\theta}_j \} &=& 0 \, \\ \nn
\EEA
Since $\theta_i$ and $\bar{\theta}_i$ are Grassmann variables, one has the following rules for the Grassmann integration 
\BEA
\int d\theta_i d\bar{\theta}_i \bar{\theta}_i \theta_i &=& 1 \\ \nn
\int d\bar{\theta}_i \theta_i &=& \int d\theta_i \bar{\theta}_i = 0 \; .
\EEA 
In our case $M_{ij} = \delta_{ij} + \tau H_{ij}$ and the determinat reads
\BEA 
\det M_{ij} &=& \int \prod_{i} d\theta_i d\bar{\theta}_{i} \;  e^{\theta_i \bar{\theta}_i + \tau \theta_i H_{ij} \bar{\theta}_j} = \\ \nn
&=& \int \prod_{i} d\theta_i d\bar{\theta}_{i} \;  e^{\theta_i \bar{\theta}_i + \tau \theta_i \frac{\partial^2 \phi }{\partial \xx_i \partial \xx_j } \bar{\theta}_j} = \\ \nn
&=& \int \prod_{i} d\theta_i d\bar{\theta}_{i} \;  e^{\theta_i \bar{\theta}_i + \tau \int d\xx d\yy \, \theta_i \delta(\xx - \xx_i) \frac{\partial^2 \phi }{\partial \xx \partial \yy } \delta(\yy - \yy_j) \bar{\theta}_j}  \; .
\EEA 
We now perform a small $\tau$ expansion that brings to
\BEA 
\det M_{ij} = \int \prod_{i} d\theta_i d\bar{\theta}_{i} \;  \left[ \theta_i \bar{\theta}_i \right. &+&  \\ \nn
\tau  \left.  \int d\xx d\yy \, \theta_i\delta(\xx - \xx_i) \frac{\partial^2 \phi }{\partial \xx \partial \yy } \delta(\yy - \yy_j) \bar{\theta}_j \right] &+& O(\tau^2) = \\ \nn
= 1 + \tau \int d\xx d\yy \, \psi( \xx ) \frac{\partial^2 \phi(x,y)}{\partial \xx \partial \yy } \psi(\yy) &+& O(\tau^2) \; .
\EEA 
Now we can write the following effective equilibrium action
\BEA \label{effective}
S_{eff}[\psi] &=&  \frac{1}{2} \int d\xx d\yy \, \psi(\xx) A(\xx,\yy) \psi(\yy)  + \\ \nn 
&&\frac{\tau}{2} \int d\xx d\yy d\zz \, \psi(\xx) \psi(\yy) \psi(\zz) B(\xx,\yy,\zz)    \\ \nn
A(x,y)      &\equiv&  \phi(\xx,\yy) - 2 \tau T \p^2_{\xx , \yy} \phi(\xx,\yy) \\ \nn
B(x,y,z)      &\equiv& \p_\zz \phi(\zz,\xx) \p_\zz \phi(\zz,\yy) \; .
\EEA
To obtain the thermodynamics of the model we have to compute the following
partition function
\BEQ
Z = \int \mathcal{D}\psi(\xx) \mathcal{D}\hat{\psi} (\xx) \, e^{-\int d\xx \, \psi(\xx) \hat{\psi} (\xx) + N \ln z - S_{eff}[\psi]}
\EEQ
where we have introduced the "single-particle" partition function $z\equiv \int d\xx \, e^{\hat{\psi}(\xx)}$.
Let us rewrite the action as follows
\BEQ
G[\psi,\hat{\psi}] \equiv  \int d\xx \, \psi(\xx) \hat{\psi} (\xx) - N \ln z + S_{eff}[\psi]
\EEQ
and thus
\BEQ
Z=\int \mathcal{D}\psi(\xx) \mathcal{D}\hat{\psi} (\xx) \, e^{- G[\psi,\hat{\psi}]}
\EEQ
The functional integral can be evaluated using the saddle-point approximation that allows to write
\BEQ 
Z \sim e^{-G[\rho,\hat{\rho}]}
\EEQ 
Where the fields $\rho(\xx)$ and $\hat{\rho}(\xx)$ satisfy the saddle-point equations
\BEA
\left.\frac{\delta G}{\delta \psi} \right|_{ \rho, \hat{\rho} } &=& \hat{\rho} + \frac{\delta S_{eff}}{\delta \rho} = 0 \\ \nn
\left. \frac{\delta G}{\delta \hat{\psi}} \right|_{\rho,\hat{\rho}} &=& \rho - \frac{N e^{\hat{\rho}} }{\int d\xx \, e^{\hat{\rho}}} = 0 \; .
\EEA
Using the first equation for eliminating $\hat{\rho}$, the free energy of the model reads
\BEA \label{lgeff}
G[\rho] &=& S_{eff}[\rho] + \int d\xx \, \rho(\xx) \ln \rho(\xx) - N \ln N \\ \nn
S_{eff}[\rho] &=& \frac{1}{2} \int d\xx d\yy \rho(\xx) A(\xx, \yy) \rho(\yy) +  \\ \nn
&& + \frac{\tau }{2} \int d\xx d\yy d\zz \rho (\xx) \rho(\yy) \rho(\zz) B(\xx,\yy,\zz)
\EEA
The action (\ref{lgeff}) reduces to the equilibrium case for $\tau \to 0$. Moreover, for $\tau \neq 0$ the theory
include a $\rho^3$ term. 
\begin{figure*}[!th]
\includegraphics[width=.75\columnwidth]{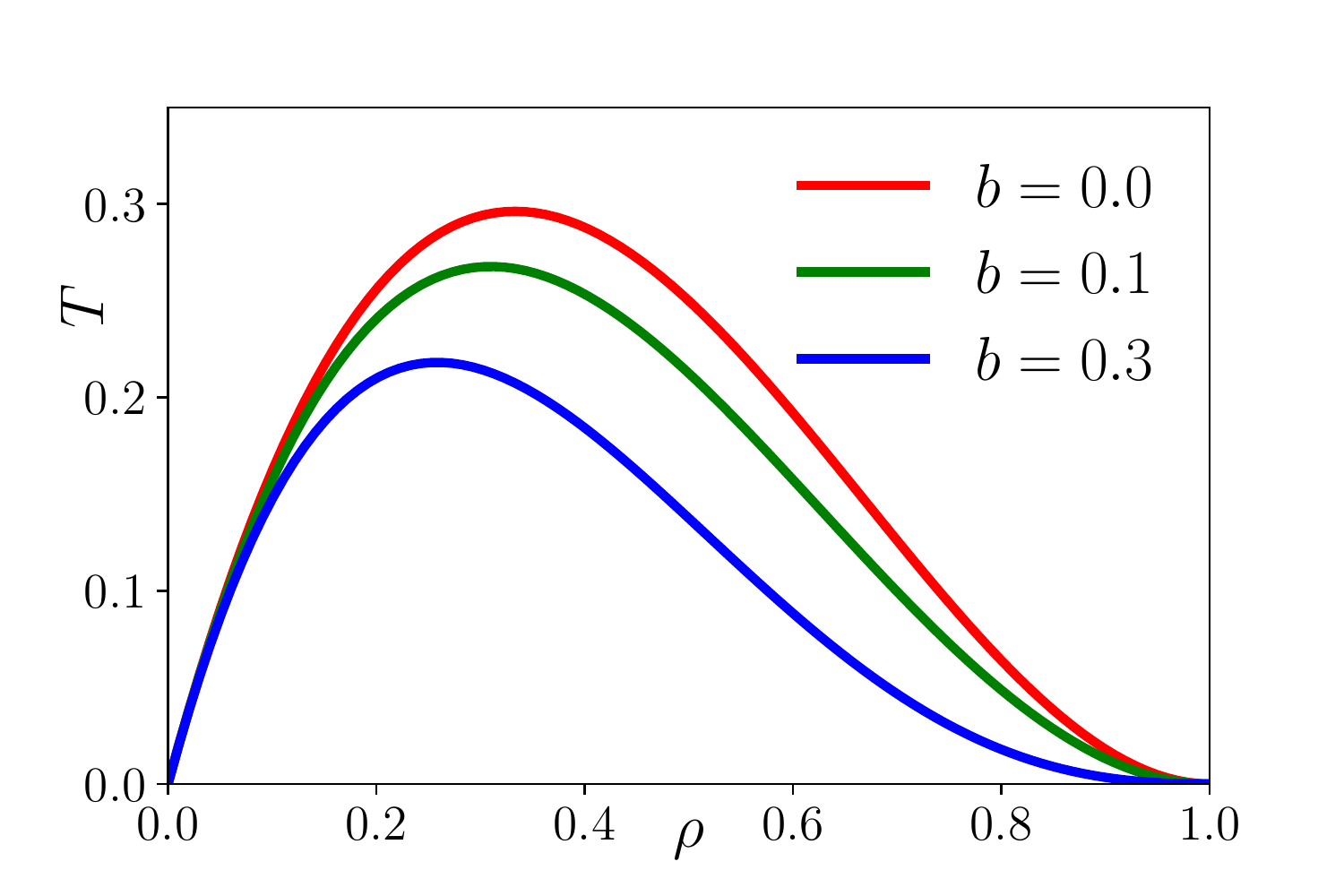}
\includegraphics[width=.75\columnwidth]{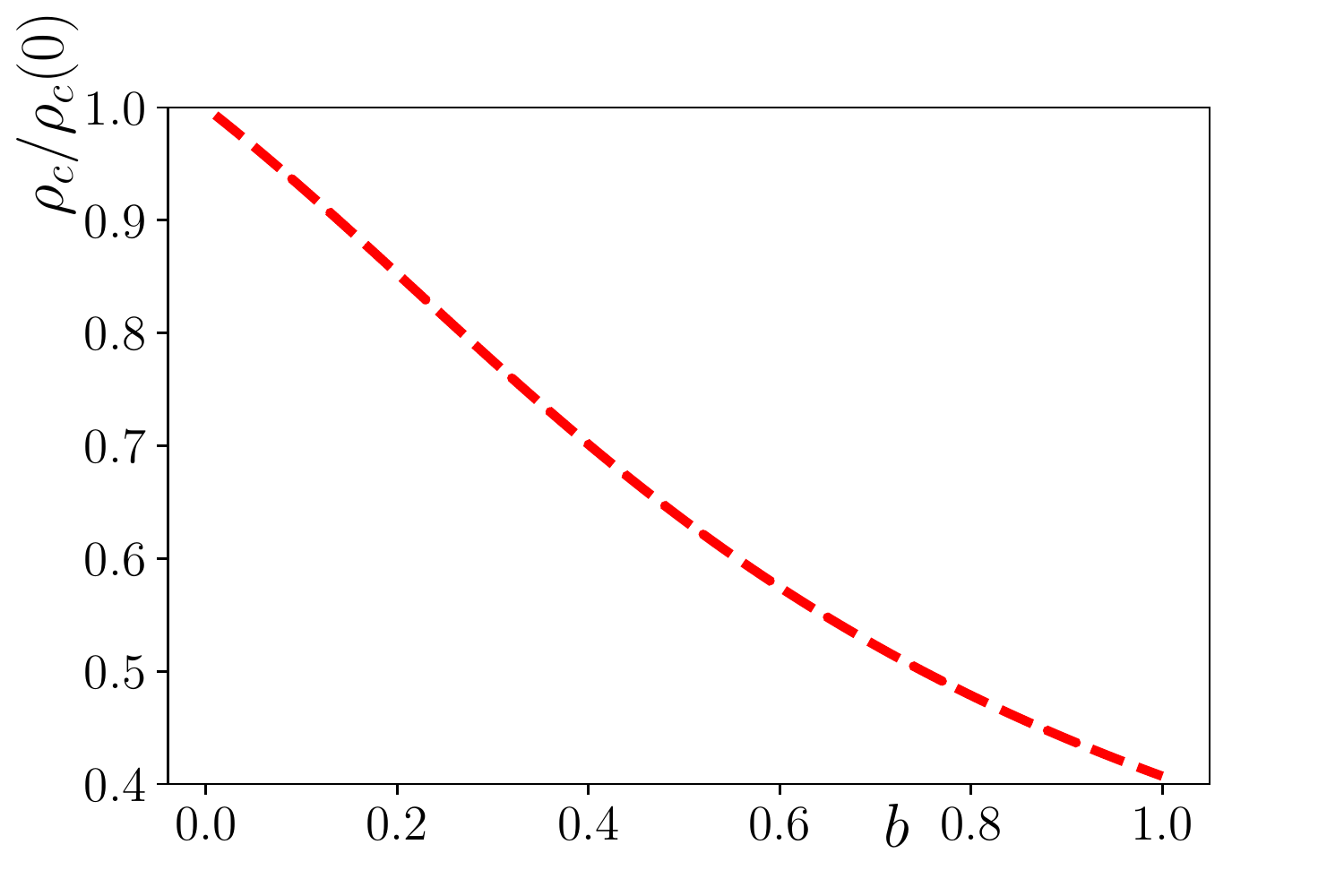}
\includegraphics[width=.75\columnwidth]{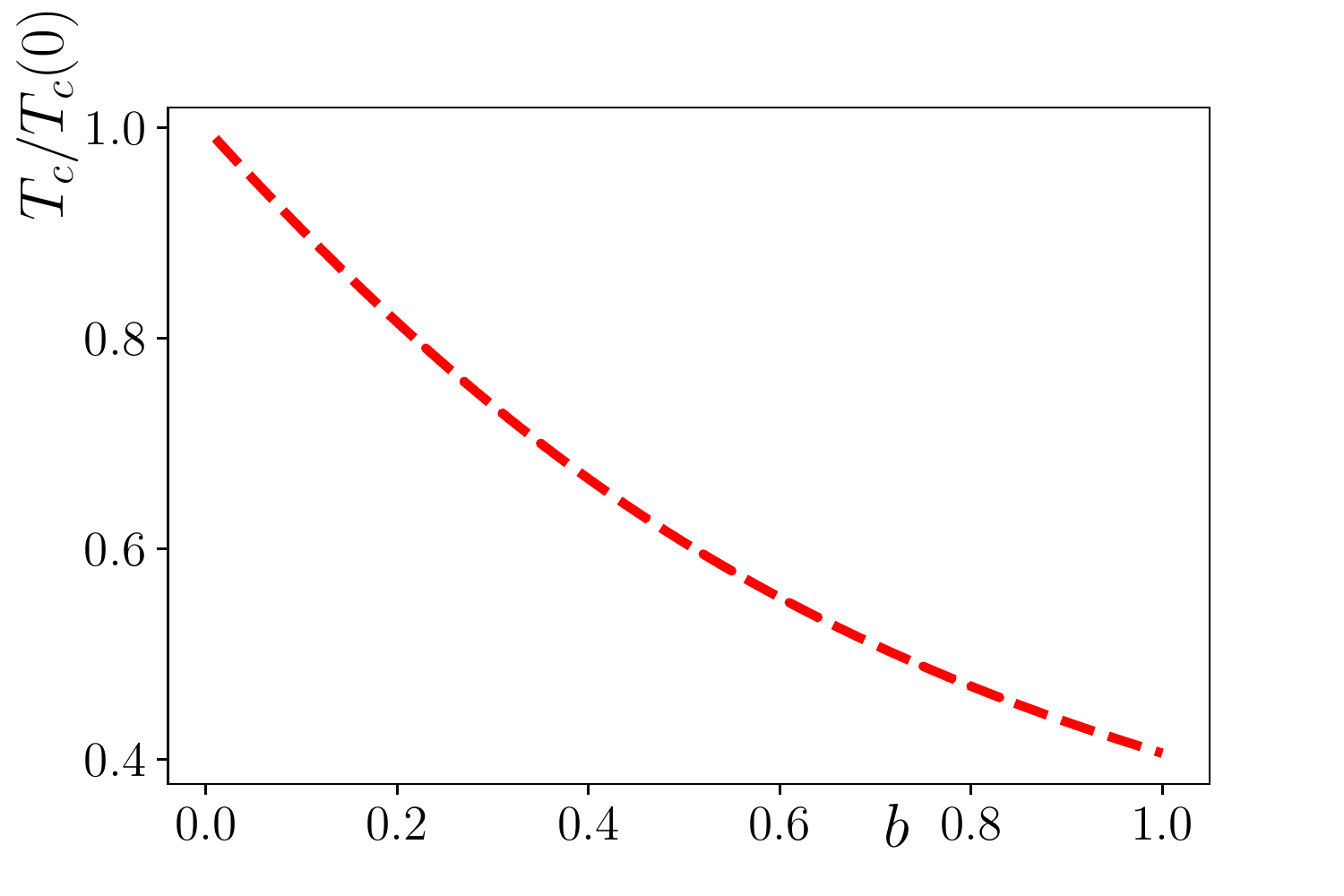}
\caption{  Phase diagram with the {\it repulsive} cubic term. Left panel: Coexistence region. Central panel: $\rho_c$ as a function of the control parameter $b$. 
Right panel: Critical temperature as a function of $b$.
\label{fig:pd_asym} 
}
\end{figure*}

\section{Computation of the effective action in the case of central pair potentials}
In the case of central pair potentials, one has
\BEQ
\phi(\xx_i,\xx_j)= \phi( |\xx_i - \xx_j | )=\phi(r_{ij})
\EEQ
where we have defined $r_{ij} \equiv |\xx_i - \xx_j | $. In $d$ spatial dimensions, 
the vector $\xx_i$ is individuated by its components $\xx_i = \{ x_i^\alpha \}_{\alpha=1}^d$
and the Euclidean distance reads
\BEQ
r_{ij}^2 = \sum_{\alpha = 1}^d (x_i^\alpha - x_j^\alpha)\; .
\EEQ
The first derivative of $\phi(r_{ij})$ with respect $\xx_i$ is
\BEQ
\frac{\p \phi(\xx_i,\xx_j)}{\p \xx_i} = \frac{1}{r_{ij}} \frac{\p \phi}{\p r_{ij}} = \frac{\phi^\prime (r_{ij}) }{r_{ij}} \; .
\EEQ
where we the prime indicates derivative with respect $r$ in $\phi(r)$.
The hessian matrix $\mathbf{H}$ of components $H_{ij}^{\alpha \beta}$ is
\BEQ
H_{ij}^{\alpha \beta} = \sum_l \left[  \frac{\p^2 \phi}{\p r_{li} } \frac{\p r_{li} }{\p x_j^\beta} \frac{\p r_{li} }{\p x_i^\alpha} + \frac{\p \phi}{\p r_{li}} \frac{\p^2 r_{li} }{\p x_i^\alpha \p x_j^\beta}  \right]
\EEQ
and the diagonal part is
\BEA
H_{ii}^{\alpha \alpha} &=& \sum_{l} \left\{  \frac{\phi^{\prime \prime} (r_{li}) }{r_{li}^2 } (x_l^\alpha - x_i^\alpha )^2 + \right. \\ \nn
&& \left. \phi^\prime (r_{li}) \left[ \frac{1}{r_{li}} -  \frac{(x_l^\alpha - x_i^\alpha )^2}{r_{li}^3 } \right] \right\} \; .
\EEA
The trace of $\mathbf{H} $ is
\BEQ
Tr \, \mathbf{H} = \sum_{i,\alpha}  H_{ii}^{\alpha \alpha} = \sum_{l,i} \left[  \phi^{\prime \prime} (r_{li}) + \frac{ \phi^\prime (r_{li})}{r_{li}} \left( d - 1 \right) \right] \; .
\EEQ
In terms of the local density field $\psi(\xx)$ we can thus write
\BEA
H_2[\psi] &=& -\tau T \int d\xx d\yy \, \psi(\xx) f( | \xx - \yy | ) \psi(\yy) \\ \nn
f(r) &\equiv& \phi^{\prime \prime}(r) + \frac{\phi^\prime (r)}{r} \left( d - 1 \right) \; .
\EEA

The term $H_1[\xx_i]$ is
\BEA
H_1[\psi] = \frac{\tau}{2} \sum_i \left( \nabla_{\xx_i} H_0 \right) \cdot  \left( \nabla_{\xx_i} H_0 \right)  = \;\;\; \;\;\;\; \;\;\;\; &&  \\ \nn
= \frac{\tau}{2} \sum_i \left( \sum_l \frac{1}{r_{il}} \frac{\p \phi }{\p r_{il} } (\xx_i - \xx_l ) \right) \cdot  \left( \sum_m \frac{1}{r_{im}} \frac{\p \phi }{\p r_{im} } (\xx_i - \xx_m ) \right) &=& \\ \nn
=\frac{\tau}{2} \sum_{i,l,m} \frac{ (\xx_i - \xx_l ) \cdot (\xx_i - \xx_m ) }{r_{il}  r_{im} }   \frac{\p \phi }{\p r_{il} } \frac{\p \phi }{\p r_{im} } = \;\;\;\;\;\; && \\ \nn
= \frac{\tau}{2} \int d\xx d\yy d\zz \, \frac{ (\xx - \yy) \cdot (\xx - \zz ) }{r_{xy} r_{xz} } \phi^\prime(r_{xy} )  \phi^\prime(r_{xz} ) \psi(\xx) \psi(\yy) \psi(\zz) && 
\EEA
At the saddle point we have $\langle \psi(\xx) \rangle = \rho(\xx)$ and thus
we can finally write the effective action for central potentials that is
\BEA
G[\rho] &=& S_{eff} [\rho] + \int d\xx \, \rho(\xx) \ln \rho(\xx)  \\ \nn
S_{eff} [\rho] &\equiv& \int d\xx d\yy \,  \rho(\xx) A( | \xx - \yy | ) \rho(\yy) + \\ \nn
&+& \frac{\tau}{2} \int d\xx d\yy d\zz \, B( |\xx - \yy| , | \xx - \zz| ) \rho(\xx) \rho(\yy) \rho(\zz)
\EEA
where the function $A(r)$ now reads
\BEQ
A(r) = \phi(r) - \tau T f(r) 
\EEQ
and the function $B(r,s)$ is
\BEQ
B(r,s) = \frac{\mathbf{r} \cdot \mathbf{s} }{r s} \phi^\prime(r) \phi^\prime(s) \; .
\EEQ

\section{Lattice models and continuum limit for $H_{UCN}$}
Here we perform the coarse-graining scheme introduced in (\ref{coarse}) in the case of $H_{UCN}$.
Again, we consider the system composed by $N$ particles each one of volume $\delta$. Particles are confined in a box of volume $V$. 
We divide the volume in small region of volume $\Delta$, each
region contains $N_i$ particles with $\sum_i N_i = N$. The partition function reads
\BEA
Z_\beta &=& \sum_{\{ N_i \} | \sum_i N_i = N} e^{-G[N_i]} \\ \nn
-G[N_i] &\equiv& \sum_{i,j} \mathcal{A}_{ij} N_i N_j + \frac{\tau}{2} \sum_{i,l,m} \mathcal{B}_{ilm} N_i N_l N_m  + \\ \nn
&& \sum_{i} \left[ N_i \ln \left( \Delta - N_i \delta \right) - N_i\ln N_i + N_i \right] \\ \nn
 \mathcal{A}_{ij} &\equiv& \beta \phi_{ij} - \tau f_{ij}\\ \nn
  \mathcal{B}_{ilm} &\equiv& \phi^\prime_{il} \phi^\prime_{im} \frac{(\rr_i - \rr_l)}{r_{il}} \cdot   \frac{(\rr_i - \rr_m)}{r_{im}}  \; .
\EEA
The continuum limit is obtained considering $\Delta \to 0$ that brings to 
\BEA
-G[\rho] &=& \int d\xx d\yy \, \mathcal{A}( |\xx - \yy| ) \rho(\xx) \rho(\yy) + \\ \nn
&+&\frac{\tau}{2} \int d\xx d\yy d\zz \, \mathcal{B}( |\xx - \yy| , |\xx - \zz|  ) \rho(\xx) \rho(\yy) \rho(\zz) + \\ \nn
&+& \int d\xx \, \left[ 1 + \ln \left( \frac{1 - \rho(\xx)}{\rho(\xx)} \right) \right]
\EEA
\begin{figure}[!th]
\includegraphics[width=1.\columnwidth]{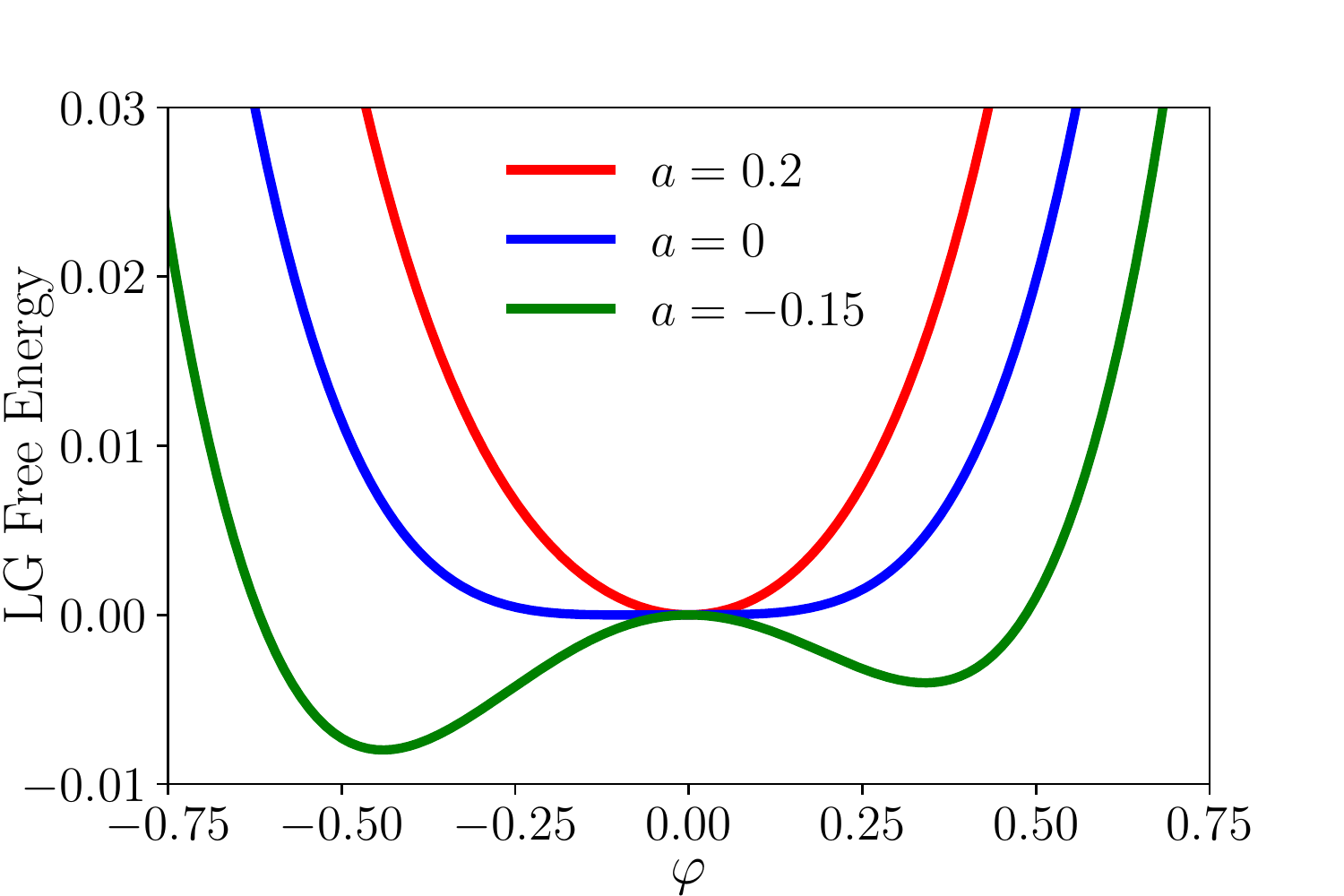} \\ 
\includegraphics[width=1.\columnwidth]{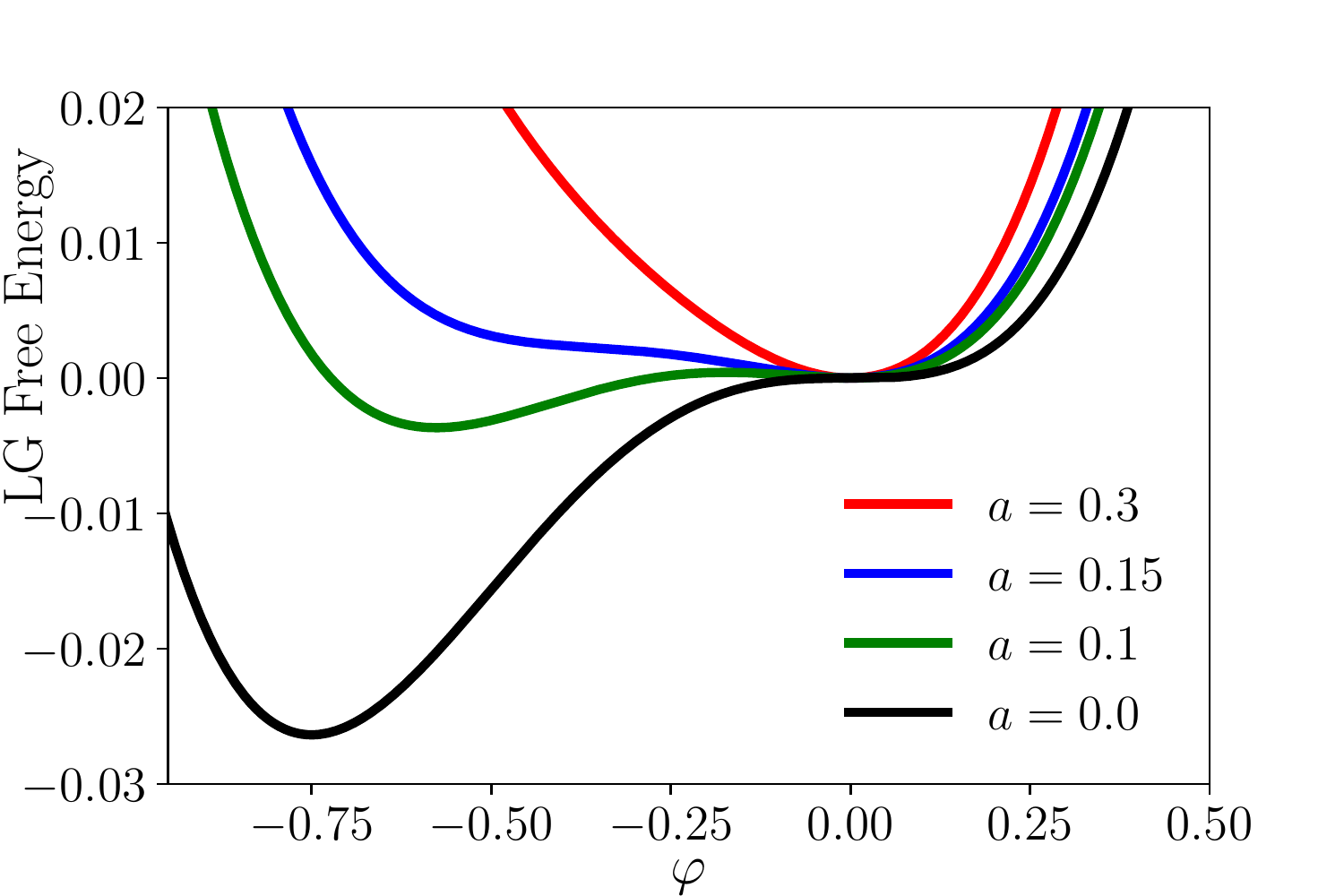}
\caption{  Landau-Ginzburg free energy (\ref{LG}) for $b^2<4 a$ at $a>0$ (upper panel) and  $b^2 \geq 4 a$ (lower panel).
\label{fig:LG_phi3} 
}
\end{figure}

\subsubsection{Mean-field theories with a cubic term}
We consider homogeneous solutions described by a real number $\rho \in [0,1]$. Now the effective action reads
\BEQ
-g(\rho) = \alpha \rho^2  + \frac{\tau b}{2}\rho^3 +  \rho \left[ \ln\frac{1-\rho}{\rho} + 1 \right] \; .
\EEQ
To make our discussion simplest possible without loss of generality, we consider only two control parameters 
that are $\alpha$ and $\tau b /2 \to b$. In particular, we are interested in evaluating the impact that the cubic
term has on spinodal decomposition. For this reason we will consider the situation $\alpha < 0$. The coexistence
region in the $T$ vs $\rho$ plane, obtained considering the solution of $\p_\rho P=0$, is
\BEQ \label{coex_b}
T(\rho,b) = 2 \rho (1 - \rho)^2 (3 b \rho - \alpha ) \, ,
\EEQ
that has been obtained considering $\alpha=-1$.
 The curves $T(\rho,b)$ are shown in Fig. (\ref{fig:pd_rep}) (left panel) for different values of the asymmetry parameter $b$. As $b$ increases we 
 obtain a contraction of the coexistence region. 
We can now compute the location of the critical point that is obtained considering the set of equations $\p_\rho P = \p_\rho^2 P = 0$.
Once we get the critical density $\rho_c(b)$, that is shown in Fig. (\ref{fig:pd_rep}), central panel, we can also compute how the critical
temperature changes with $b$ (same figure, right panel). It turns that both, the critical density and the critical temperature, are decreasing
function of $b$.

\subsubsection{Landau-Ginzburg $\varphi^3$ Theory} \label{LGPHI3}
In this section, we provide a brief discussion about $\varphi^3$ field theories in Statistical Physics. In particular, we consider both 
interactions, $\varphi^3$ and $\varphi^4$ in order to have a well-defined Landau-Ginzburg energy functional.
For $b>0$, the critical behavior of the system can be studied through the following Landau-Ginzburg free energy
\BEQ \label{LG}
F_{LG}[\varphi] = \int d\xx \left[ \frac{1}{2} (\nabla \varphi)^2 + \frac{a}{2} \varphi(\xx)^2  + \frac{b}{3} \varphi(\xx)^3 + \frac{c}{4} \varphi(\xx)^4 \right]
\EEQ 
where the order parameter $\varphi(\xx)$ represents fluctuations around the critical density $\rho_c$, i. e., $\rho(\xx) = \rho_c + \varphi(\xx)$.
Without loss of generality, we consider the homogeneous case $\varphi(\xx)=\varphi$. We immediately realize that the cubic term, breaking 
the symmetry $\varphi \to -\varphi$, promotes one phase with respect to the other \cite{barrat2003basic}.
Minimizing Eq. (\ref{LG}), one obtains
three configurations $\varphi_{0,1,2}$ that are
\BEA
\varphi_0 &=& 0 \\ \nn
\varphi_{1,2} &=& \frac{- b}{2 \, c} \pm \frac{1}{2\, c} \sqrt{    b^2 - 4 \, a \, c } \, .
\EEA
For sake of simplicity, let us fix $c=1$. We consider the coefficients $a$ and $b$ as two independent and tunable external parameters of our
coarse-grained model. For $b=0$, the coefficient $a$ changes sign at the MIPS critical point. For $b \neq 0$ the value $a=0$ is not necessary the MIPS critical point. 
In particular, the coefficient $b$ tunes the intensity of asymmetry $\varphi \to -\varphi$. 
For $b=0$, we recover the standard $\varphi^4$ theory, and thus the symmetry $\varphi \to - \varphi$ is preserved. For $b>0$ and $b^2<4 a$, at $a>0$ the only real and stable solution is 
$\varphi_0$ that becomes marginal at $a=0$ and eventually unstable for $a<0$. In the latter case, $\varphi_{1,2}$ become reals and the solution $\varphi_1$
is the new minimum. When $b^2 \geq 4 a$,  $\varphi_{1,2}$ are reals for $a>0$, meaning that the system develops a metastable state. The two situations are showed in
Fig. (\ref{fig:LG_phi3}).

\bibliography{mpbib}

\end{document}